\documentclass[%
 reprint,
nofootinbib,
 amsmath,amssymb,
 aps,
prb,
]{revtex4-2}

\usepackage{footnote}
\usepackage{multirow}
\usepackage{graphicx}
\usepackage{subfigure} 
\usepackage{dcolumn}
\usepackage{bm}
\usepackage[compat=1.1.0]{tikz-feynhand}
 \usepackage{bbm}
 \usepackage[colorlinks,linkcolor=red,anchorcolor=blue,citecolor=blue]{hyperref}

\newcommand{\tabincell}[2]{\begin{tabular}{@{}#1@{}}#2\end{tabular}}

\newcommand{\jpsi}{J/\psi}
\newcommand{\chiczero}{\chi_{c0}}
\newcommand{\chicone}{\chi_{c1}}
\newcommand{\chictwo}{\chi_{c2}}
\newcommand{\Bp}{B^{+}}
\newcommand{\Dp}{D^{+}}
\newcommand{\Dm}{D^{-}}
\newcommand{\Dsp}{D^{+}_{s}}
\newcommand{\Dsm}{D^{-}_{s}}
\newcommand{\Kp}{K^{+}}

\begin{document}

\preprint{APS/123-QED}

\title{On the nature of $X(3960)$}

\author{Yu Chen}
\affiliation{Department of Physics and State Key Laboratory of Nuclear Physics and Technology, Peking University, Beijing 100871, China}
\author{Hao Chen}
\affiliation{Department of Physics and State Key Laboratory of Nuclear Physics and Technology, Peking University, Beijing 100871, China}
\author{Ce Meng}
\affiliation{Department of Physics and State Key Laboratory of Nuclear Physics and Technology, Peking University, Beijing 100871, China}
\author{Hong-Rong Qi}
\thanks{Corresponding author: hrqi@ihep.ac.cn}
\affiliation{Department of Physics, National Taiwan University, Taipei 10617, Chinese Taiwan}
\author{Han-Qing Zheng}
\thanks{Other e-mail addresses: yuyuyu@pku.edu.cn (Yu Chen), haochen0393@pku.edu.cn (Hao Chen), mengce75@pku.edu.cn (Ce Meng), zhenghq@scu.edu.cn (Han-Qing Zheng).}
\affiliation{College of Physics, Sichuan University, Chengdu, Sichuan 610065, China}

\date{\today}

\begin{abstract}
A near-threshold enhancement in the $\Dsp\Dsm$ system, dubbed as $X(3960)$, is observed by the LHCb collaboration recently.
A combined analysis on $\chiczero(3930)~(\to D^+ D^-)$, $X(3960)~(\to\Dsp\Dsm)$, and
$X(3915)~(\to\jpsi\omega)$ is performed using both a $K$-matrix approach of $D_{(s)}\bar{D}_{(s)}$ four-point contact interactions and a model of  Flatt\'e-like parameterizations. The use of the pole counting rule and  spectral density function sum rule 
indicate, under current statistics, that this $\Dsp\Dsm$ near-threshold state has probably 
the mixed nature of  a $c\bar{c}$ confining state and $\Dsp\Dsm$ continuum.  
\end{abstract}
\pacs{12.40-y,~13.25.Hw,~13.75.Lb}
\maketitle


\section{\label{sec:level1}
Introduction}

Most recently, the LHCb experiment~\cite{LHCb-PAPER-2022-019,LHCb-PAPER-2022-018} announced a new hadron, $X(3960)$, observed in the $B^+ \to \Dsp\Dsm K^+$ process, where the properties of this state are measured to be
\begin{align}
 M~~~       &= 3956 \pm 5 \pm 10 \,{\rm MeV}, \nonumber\\
 \Gamma~~~~ &= 43\pm 13\pm 8 \,{\rm MeV}, \nonumber\\
 J^{PC}     &=0^{++}.\nonumber
\end{align}
If taking the $X(3960)$ state and the $\chiczero(3930)$ in the $D^+D^-$ mass distribution~\cite{LHCb-PAPER-2020-025} as the same particle, the partial-width ratio~\cite{LHCb-PAPER-2022-018} is calculated to be
\begin{eqnarray}
\frac{\Gamma(X\to D^+D^-)}{\Gamma(X\to D_s^+D_s^-)}&=&0.29\pm0.09\pm0.10\pm0.08, \nonumber
\end{eqnarray}
which probably implies the exotic nature of this hadron, since it is harder to excite an $s\bar{s}$ pair than $u\bar{u}$ or $d\bar{d}$ pairs from the vacuum if this state is a pure charmonium~\cite{LHCb-PAPER-2022-018}.
Several recent theoretical models~\cite{Bayar:2022dqa,Ji:2022vdj,Xin:2022bzt,Chen:2022dad,Mutuk:2022ckn} believe that the $X(3960)$ state is a molecular $\Dsp\Dsm$ structure, while others take it as a  scalar $[cs][\bar{c}\bar{s}]$ tetraquark~\cite{Agaev:2022pis,Guo:2022crh,Badalian:2023qyi}.

Besides the $X(3960)$ and $\chiczero(3930)$ states, the $X(3915)$ found in the $\jpsi\omega$ mass spectrum~\cite{PDG2022} is also near the $\Dsp\Dsm$ threshold.
As the three states have compatible masses and widths, as well as the preferred $J^{PC}=0^{++}$ assignment~\cite{PDG2022}, we assume that they are the same hadron, noted $X$, in this work.
A combined analysis on the nature for this $X$ state is performed 
using both a model of $D\bar{D}$ and $D_s^+D_s^-$ four-point contact interactions and an energy-dependent Flatt\'e-like parameterization. 
The pole counting rule (PCR)~\cite{Morgan:1992ge}, which has been generally applied to the studies of ``$XYZ$" physics in Refs.~\cite{Zhang:2009bv,Dai:2012pb,Cao:2019wwt,Chen:2021tad,Gong:2016hlt,Cao:2020gul}, and spectral density function sum rule (SDFSR)~\cite{Gong:2016hlt,Cao:2020gul,Baru:2003qq,Weinberg:1962hj,Weinberg:1965zz,Kalashnikova:2009gt} (only available for a Flatt\'e(-like) model) are employed to distinguish whether this $X$ state is more inclined to be confining state bound by color force, or composite hadronic molecule loosely bound by deuteron-like meson-exchange force.

In this paper, Section~\ref{sec:kmatrix} presents a couple-channel $K$-matrix approach without the explicitly-introduced $X$ state~\footnote{Explicit $X$ state means that a field of $X$ is introduced in the lagrangians.}  to model the $D_{(s)}\bar{D}_{(s)}$ rescattering, Section~\ref{sec:flatte} exploits a parameterization with explicitly-introduced $X$ state, using an energy-dependent Flatt\'e-like formula, to describe the couplings of $X\to D^+D^-, \Dsp\Dsm$, and  $\jpsi\omega$, and a brief summary and discussion are carried out in the last section.

\section{$K$-matrix Approach}
\label{sec:kmatrix}

For a comprehensive study of this $X$ state's nature, four decays, $i.e.$ $B^+\to D^+D^-K^+$, $\Bp\to\Dsp\Dsm\Kp$, $B^+\to \jpsi\omega K^+$, and $\gamma^*\gamma^*\to\jpsi\omega$, are utilized in this article.
In this section, a couple-channel $K$-matrix approach with the implicitly-introduced $X$ state is employed to get the unitarized amplitudes. Then, poles of unitarized amplitudes will be searched for in the complex $s$ planes, and PCR~\cite{Morgan:1992ge} is implemented to investigate the nature of this $X$ structure near the $\Dsp\Dsm$ threshold.

\subsection{Amplitudes of $D_{(s)}\bar{D}_{(s)}$ rescattering effect }
The effective lagrangians for the above four channels under $S$-wave couplings can be constructed as follows:

\begin{equation}
    \begin{aligned}
\mathcal{L}_{B^+D\bar{D}K^+}&=g_{B1} B^+D\bar{D}K^+\ ,  \\
\mathcal{L}_{B^+D_s^+ D_s^-K^+}&=g_{B 2} B^+D_s^+ D_s^-K^+\ ,\\
\mathcal{L}_{\gamma^* \gamma^* D \bar{D}}&=g_{\gamma 1} F^{\mu \nu}F_{\mu \nu}D \bar{D}\ ,\\
\mathcal{L}_{\gamma^* \gamma^* D_s^+ D_s^-}&=g_{\gamma 2} F^{\mu \nu}F_{\mu \nu}D_s^+ D_s^-\ ,\\
\mathcal{L}_{D\bar{D}D\bar{D}}&=g_{11}D\bar{D}D\bar{D}\ ,\\
\mathcal{L}_{D\bar{D} D_s^+ D_s^-}&=g_{12} D\bar{D} D_s^+ D_s^-\ ,\\
\mathcal{L}_{D\bar{D} J/\psi\omega}&= g_{13} D\bar{D} \psi_{\mu}\omega^{\mu}\ ,\\
\mathcal{L}_{D_s^+ D_s^-D_s^+ D_s^-}&=g_{22} D_s^+ D_s^-D_s^+ D_s^-\ ,\\
\mathcal{L}_{D_s^+ D_s^- J/\psi\omega}&= g_{23} D_s^+ D_s^- \psi_{\mu}\omega^{\mu}\ ,\\
    \end{aligned}  \label{lagrangian}
\end{equation}
where $D=(D^0,D^+)^T$, $\bar{D}=(\bar{D}^0,D^-)$, $g$ with  subscripts stand for coupling constants, and the subscripts 1, 2 and 3 refer to the $D\bar{D}$, $D_s^+D_s^-$ and $J/\psi \omega$ channels, respectively. For example, $g_{B1}$  means the coupling constant of $B^+ \rightarrow D \bar{D} K^+$ four-point vertex.

In this work, $D\bar{D}$ will be written in isospin eigenstate, $i.e.$ $|D\bar{D}\rangle^{I=0}=\frac{-1}{\sqrt{2}}(|D^0 \bar{D}^0\rangle+|D^+ D^-\rangle)$ and $|D\bar{D}\rangle^{I=1}=\frac{1}{\sqrt{2}}(|D^0 \bar{D}^0\rangle-|D^+ D^-\rangle)$~\footnote{Note that the isospin state of $|D^+\rangle$ is $-|I,I_3\rangle=-|\frac{1}{2},\frac{1}{2}\rangle$, $|D^-\rangle=|\frac{1}{2},\frac{-1}{2}\rangle$, $|D^0\rangle=|\frac{1}{2},\frac{-1}{2}\rangle$,   and $|\bar{D}^0\rangle=|\frac{1}{2},\frac{1}{2}\rangle$, where these conventions can guarantee $G$ and $C$ parities are conserved~\cite{Du:phd}.}. Two channels, $D\bar{D}$ and $D_s^{+}D_s^{-}$, are considered to construct the $K$-matrix. From the lagrangians in Eq.~(\ref{lagrangian}), the $K$-matrix can be written as
\begin{equation}
\mathbf{K}^{I=0}=\left(\begin{array}{cc}
6g_{11} & -\sqrt{2}g_{12} \\
-\sqrt{2}g_{12} & 4g_{22}
\end{array}\right)\ , \label{KI=0}
\end{equation}
in which $\mathbf{K}_{11}$ stands for the process $D\bar{D}\to D\bar{D}$, and $\mathbf{K}_{12}$ for $D\bar{D}\to D_s^{+}D_s^{-}$, etc. The unitarized amplitudes can be obtained by
\begin{equation}
   \mathbf{T}=\mathbf{K}\cdot [1-\mathbf{G}\mathbf{K}]^{-1}\ ,
\end{equation}
where $\mathbf{G}$ is the diagonal matrix of two-point loop integrals of these two channels, $i.e.$ $\mathbf{G}=\mathrm{diag}\left[B_0(m_{D},m_{\bar{D}}), B_0(m_{D_s^+},m_{D_s^-})\right]$. The definition of two-point loop integral is,
\begin{equation}
\begin{aligned}
    B_{0}&(p^2, m_1,m_2)=\\
    &\frac{\mu^{\epsilon}}{i}\int \frac{d^D k}{(2 \pi)^D}\frac{i}{k^2- m_1^2+i0^+}\frac{i}{(p-k)^2-m_2^2+i 0^+}\\
    \sim & \frac{1}{16\pi^2}\left[-R+1-\text{ln}\frac{m_1^2}{\mu^2}+\frac{m_1^2-m_2^2-p^2}{2p^2}\text{ln}\frac{m_2^2}{m_1^2} \right.\\
    &\left. +\frac{p^2-(m_1-m_2)^2}{p^2}\alpha(p^2)\text{ln}\frac{\alpha(p^2)-1}{\alpha(p^2)+1}\right]
    \ ,\label{B0 function}
\end{aligned}
 \end{equation}
 where a $\overline{MS}$ renormalization is understood to be taken, 
 $\alpha(p^2)\equiv \sqrt{\frac{p^2-(m_1+m_2)^2}{p^2-(m_1-m_2)^2}}$, $\mu$ is renormalization scale set to be 1~GeV, $a(\mu)=-R+1$ is the subtraction constant as a parameter which is $8.05$ with large uncertainties obtained by the fit~\footnote{In Ref.~\cite{Oller:2000fj}, $a(\mu)$ is taken to be $\simeq2$. There is no physics here, however. If we take $a(\mu)=2$, the fit gives $\chi^2/d.o.f.=56.29/51=1.10$, with the pole position almost unchanged, $i.e.~\sqrt{s}=(3.9192 - 0.0130 i)~\text{GeV}$.}.

Since $X(3960)$ has isospin $I=0$, only the iso-singlet states, $|D\bar{D}\rangle^{I=0}$ and $|D_s^+ D_s^-\rangle$, are included in the unitarized amplitudes $\mathbf{T}$. But the experimental results have been reported only in the $D^+D^-$ final states to date, so the approximations
\begin{equation}
\begin{aligned}
        T_{D^0 \bar{D}^0\to D^+D^-}=\frac{1}{2}(T^{I=0}-T^{I=1})\sim \frac{1}{2}T^{I=0}\ ,\\
        T_{D^+ D^-\to D^+ D^-} = \frac{1}{2}(T^{I=0}+T^{I=1})\sim \frac{1}{2}T^{I=0}\ ,
\end{aligned} 
\end{equation}
are used. With the help of $\mathbf{T}$, the amplitude of $B^+\to D^+ D^- K^+$ can be written as
\begin{equation}
     \mathcal{A}_{B^+\to D^+D^-K^+} = g_{B1}\mathbf{T}_{11}-\frac{1}{\sqrt{2}}g_{B2}\mathbf{T}_{21}\ ,
\end{equation}
which satisfies the constraint on final-state interactions~\cite{pennington}~\footnote{Our parameterization is actually equivalent to another widely used form, $e.g.$, in Ref.~\cite{Zhu:2022guw}, up to a mild background contribution.}.

 Other amplitudes of $B^+\to D_s^+D_s^-K^+,~ B^+\to J/\psi\omega K^+,~\gamma^*\gamma^*\to J/\psi \omega$, in which the interactions of intermediate $D\bar{D}$ and $D_s^+D_s^-$ states  are included, can be expressed as 
\begin{equation}
    \mathcal{A}_{B^+\to D_s^+D_s^- K^+} = -\sqrt{2}g_{B1}\mathbf{T}_{12}+g_{B2}\mathbf{T}_{22}\ , 
\end{equation}

\begin{equation}\label{B_decay_psi_omega}
    \begin{aligned}
        \mathcal{A}_{B^+\to J/\psi \omega K^+}& = \epsilon^*(p_{\psi})\cdot\epsilon^*(p_{\omega})\times\left(2g_{B1}\mathbf{T}_{11}g_{13}\right. \\
        &\left.-\sqrt{2}g_{B1}\mathbf{T}_{12}g_{23}-\sqrt{2}g_{B2}\mathbf{T}_{21}g_{13}+g_{B2}\mathbf{T}_{22}g_{23}\right)\ ,
    \end{aligned}
\end{equation}
\begin{equation}\label{double_gamma}
    \begin{aligned}
        \mathcal{A}_{\gamma^*\gamma^*\to J/\psi \omega 
  }=&\left(p_{\gamma_1}^{\mu}p_{\gamma_2}^{\nu}-p_{\gamma_1}\cdot p_{\gamma_2}g^{\mu\nu}\right)\epsilon_{\mu}(p_{\gamma_1})\epsilon_{\nu}(p_{\gamma_2})\\
        &\times\left(8g_{\gamma 1}\mathbf{T}_{11}g_{13}-4\sqrt{2}g_{\gamma 1}\mathbf{T}_{12}g_{23}\right.\\
        &\left.-4\sqrt{2}g_{\gamma 2}\mathbf{T}_{21}g_{13}+4g_{\gamma 2}\mathbf{T}_{22}g_{23}\right)\\
        &\times \epsilon^{*}(p_{\psi})\cdot \epsilon^{*}(p_{\omega})\ .
    \end{aligned}
\end{equation}
Note that in Eqs.~(\ref{B_decay_psi_omega}) and (\ref{double_gamma}), the final processes $D_{(s)}\bar{D}_{(s)}\to J/\psi \omega$ are included by multiplying corresponding tree-level diagrams, 
as it is suppressed by the Okubo-Zweig-Iizuka (OZI) rule.


\subsection{Numerical results and pole analysis}
In the previous subsection, the amplitude involving this $X$ state in each decay is obtained. Now a simultaneous analysis for the aforementioned four  processes is performed to  fit the experimental data~\cite{LHCb-PAPER-2020-025,LHCb-PAPER-2022-018,LHCb-PAPER-2022-019,BaBar:2010wfc,BaBar:2012nxg}.
The background (BKG) shapes are parameterized to be similar as those in the experiments.
For the $\Dp\Dm$ chain, the incoherent background contains two charmonia, $\psi(3770)$ and $\chictwo(3930)$, modelled by Breit-Wigner functions,  and the mass reflection of the $X_1(2900)$ resonance described by a 1st-order polynomial times a Gaussian $\mathcal{G}(\mu,\sigma)$ where the parameters are extracted by fitting to the $X_1(2900)$ component in the LHCb data~\cite{LHCb-PAPER-2020-025}; and the three-body phase space of $\Bp\to\Dp\Dm K^+$ is applied to describe other potential coherent backgrounds with unconsidered intermediate states. 
For the $\Dsp\Dsm$ mode, the backgrounds below 4.25 GeV in the invariant $\Dsp\Dsm$ mass are the $X_0(4140)$ state and the non-resonant three-body phase space of $\Bp\to\Dsp\Dsm K^+$, which are coherent with the $X$ state on grounds of the LHCb analysis~\cite{LHCb-PAPER-2022-018}.
For the two $\jpsi\omega$ decays, only incoherent backgrounds are taken into account, which is in agreement with the experiments~\cite{BaBar:2010wfc,BaBar:2012nxg}.

Finally, the number of events for each decay can be expressed by~\cite{Hanhart:2007yq} 
\begin{equation}
    \begin{aligned}
        &N_{B^+\rightarrow D^+D^-K^+}(s)\\
        &=\frac{\mathcal{R}_{D^+D^-}\mathcal{N}_{X\to D^+D^-}}{\mathcal{B}(B^+ \rightarrow XK^+)\mathcal{B}(X\rightarrow D^+D^-)\Gamma_B} \frac{1}{(2 \pi)^3 32 m_B} \\
        & \ \ \times 2 \sqrt{s} \rho(m_B,m_K,\sqrt{s}) \rho(\sqrt{s},m_D,m_D) \\
        & \ \ \times |\mathcal{A}_{B^+ \rightarrow D^+D^- K^+}+a_1 e^{i \phi_1}|^2\\
                 & \ \ +\left|\frac{g_{\psi(3770)}}{s-m_{\psi(3770)}^2+i m_{\psi(3770)} \Gamma_{\psi(3770)}}\right|^2 \\
        & \ \ +\left|\frac{g_{\chictwo(3930)}}{s-m_{\chictwo(3930)}^2+i m_{\chictwo(3930)} \Gamma_{\chictwo(3930)}}\right|^2 \\
        & \ \ +a \sqrt{s} e^{\frac{-(\sqrt{s}-\mu)^2}{2 \sigma^2}} \ ,
    \end{aligned} \label{Nbkdd1}
\end{equation}

\begin{figure*}[htb]
\centering
\includegraphics[width=0.495\textwidth]{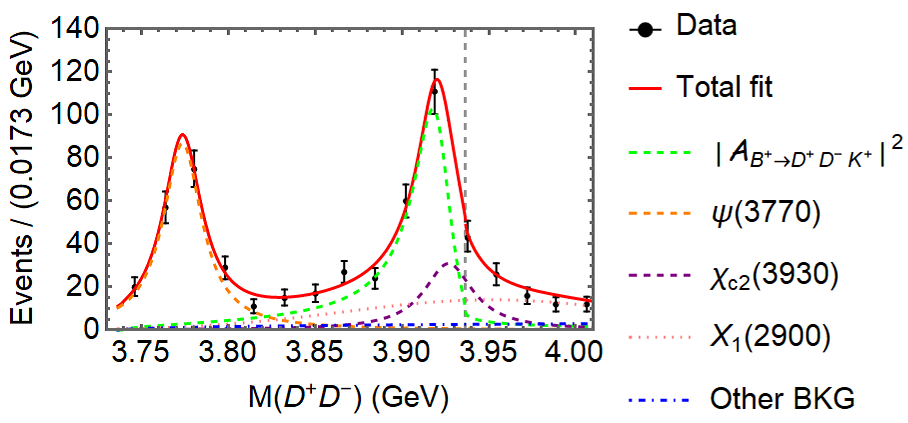}
\includegraphics[width=0.495\textwidth]{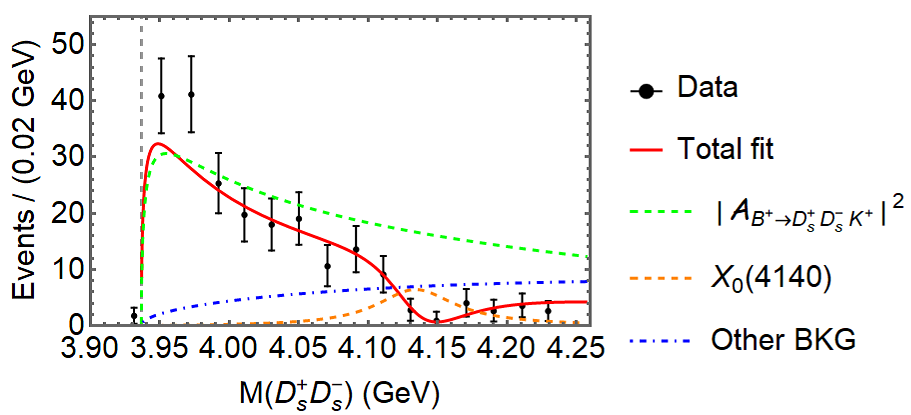} \\
\vspace{0.2cm}
\includegraphics[width=0.495\textwidth]{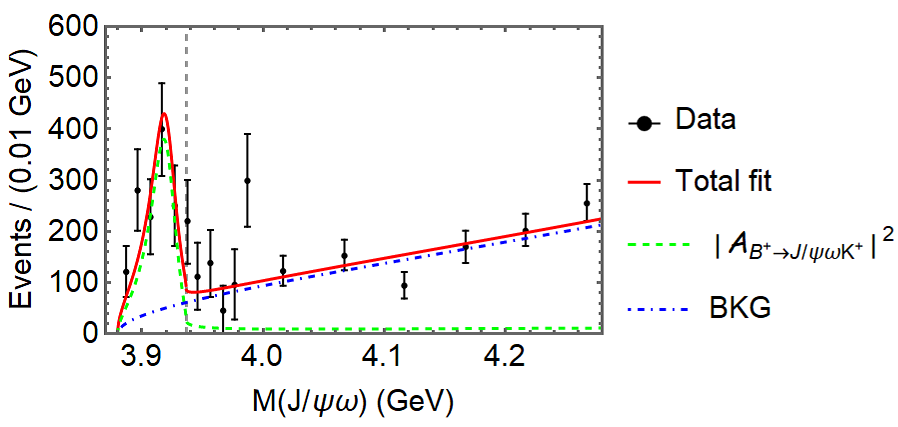}
\includegraphics[width=0.495\textwidth]{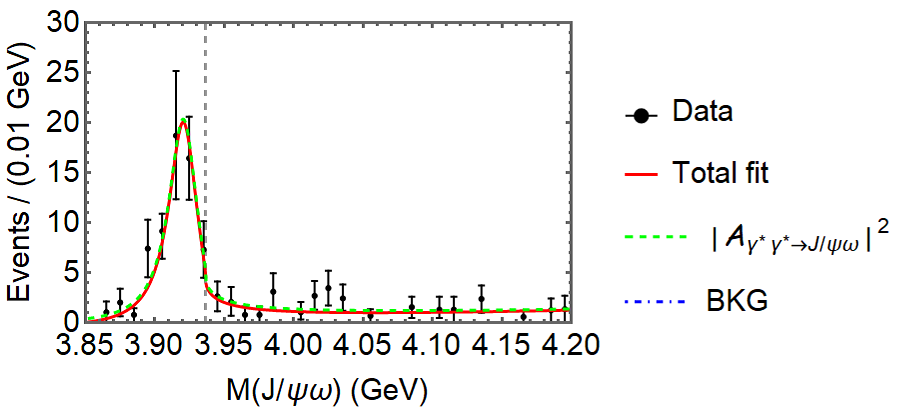}

\caption{(Color online) Fit distributions with the $K$-matrix approach for $B^+ \to \Dp\Dm\Kp$ (top left), $B^+ \to D_s^+ D_s^-\Kp$ (top right), $B^+ \to \jpsi \omega \Kp$ (bottom left), and $\gamma^*\gamma^* \to \jpsi \omega$ (bottom right). Here, the data are from Refs.~\cite{LHCb-PAPER-2020-025,LHCb-PAPER-2022-018,LHCb-PAPER-2022-019,BaBar:2010wfc,BaBar:2012nxg}, and the vertical dashed lines are located at the $D_s^+ D_s^-$ threshold. }
\label{fig:bubble2}
\end{figure*} 
\begin{figure*}
    \centering
    \includegraphics[width=17cm]{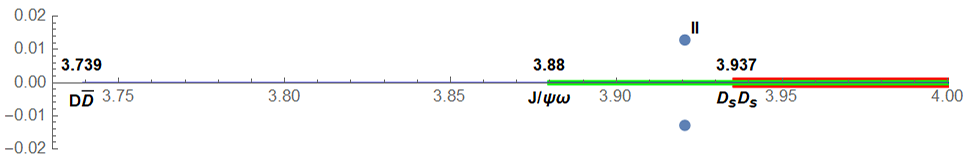}
    \caption{(Color online) Schematic diagram of pole positions for the $K$-matrix approach.}
    \label{fig:pole_bubble2}
\end{figure*}

\begin{equation}
    \begin{aligned}
        &N_{B^+ \rightarrow D_s^+ D_s^- K^+}(s)\\
        &=\frac{\mathcal{R}_{D_s^+D_s^-}\mathcal{N}_{X\to D_s^+D_s^-}}{\mathcal{B}(B^+\rightarrow XK^+)\mathcal{B}(X\rightarrow D_s^+D_s^-)\Gamma_B}\frac{1}{(2 \pi)^3 32m_B} \\
        & \ \ \times 2 \sqrt{s} \rho(m_B,m_K,\sqrt{s})\rho(\sqrt{s},m_{D_s},m_{D_s})\\
        & \ \ \times \left| \mathcal{A}_{B\rightarrow  D_s^+ D_s^- K^+} +a_2 e^{i \phi_{21}} \right.\\
        &\left. \ \ +\frac{g_{X_0(4140)} e^{i \phi_{22}}}{s-m_{X_0(4140)}^2+i m_{X_0(4140)}\Gamma_{X_0(4140)}}\right|^2\ ,
    \end{aligned} \label{Ndsds1}
\end{equation}
\begin{equation}
    \begin{aligned}
             &N_{B^+\rightarrow J/\psi \omega K^+}(s)\\
             &=\frac{\mathcal{R}_{J/\psi \omega}\mathcal{N}_{X\to\jpsi\omega}}{\mathcal{B}(B^+\rightarrow X K^+)\mathcal{B}(X \rightarrow J/\psi \omega)\Gamma_B} \frac{1}{(2 \pi)^3 32 m_B}\\
        & \ \ \times 2 \sqrt{s}\rho(m_B,m_K,\sqrt{s}) \rho(\sqrt{s},m_{J/\psi},m_{\omega})\\
        & \ \ \times  
             |\mathcal{A}_{B^+ \rightarrow J/\psi \omega K^+}|^2\\
            & \ \ +a_{31} \rho(\sqrt{s},m_{J/\psi},m_{\omega})e^{a_{32}(\sqrt{s}-m_{J/\psi}-m_{\omega})} \ ,
    \end{aligned} \label{Nbkpo1}
\end{equation}
\begin{equation}
\begin{aligned}
   N_{\gamma^* \gamma^* \rightarrow J/\psi \omega}(s)&=n_4 \rho(\sqrt{s},m_{J/\psi},m_{\omega})| \mathcal{A}_{\gamma^* \gamma^* \rightarrow J/\psi \omega}|^2\\
        & \ \ +(a_{41} \sqrt{s}+a_{42})\ ,  
\end{aligned}
   \label{Ngamma1}
\end{equation}
where $\mathcal{R}_{\Dp\Dm}=0.0173$ GeV, $\mathcal{R}_{D_s^+D_s^-}=0.02$ GeV, and $\mathcal{R}_{J/\psi \omega}=0.01$ GeV, are intervals of invariant mass spectra in the experimental data; 
$\mathcal{N}_{X\to\Dp\Dm}$, $\mathcal{N}_{X\to D_s^+D_s^-}$, and $\mathcal{N}_{X\to\jpsi\omega}$, are the numbers of the expected $X$ signal events in the $\Bp\to\Dp\Dm\Kp$, $\Dsp\Dsm\Kp$, and $\jpsi\omega\Kp$ decays, respectively, with $\mathcal{N}_{X\to\Dp\Dm}=46.6$ and $\mathcal{N}_{X\to D_s^+D_s^-}=91.4$ obtained by experiments~\cite{LHCb-PAPER-2020-025,LHCb-PAPER-2022-018}, while $\mathcal{N}_{X\to\jpsi\omega}$ is an unkown parameter to be fitted; the decay branching fractions $\mathcal{B}(B^+ \rightarrow XK^+)\mathcal{B}(X\rightarrow D^+D^-)=8.1\times 10^{-6}$~\cite{Zyla:2020zbs}, $\mathcal{B}(B^+\rightarrow XK^+)\mathcal{B}(X\rightarrow D_s^+D_s^-)=\mathcal{B}(B^+ \rightarrow XK^+)\mathcal{B}(X\rightarrow D^+D^-)\frac{\Gamma(X\to D_s^+D_s^-)}{\Gamma(X\to D^+D^-)}=\frac{8.1 \times 10^{-6}}{0.29}$, and $\mathcal{B}(B^+\rightarrow X K^+)\mathcal{B}(X \rightarrow J/\psi \omega)=3.0 \times 10^{-5}$~\cite{BaBar:2010wfc};
$\Gamma_B=4.018\times 10^{-10}$ MeV, is the width of the $\Bp$ meson in accordance  with the relationship $\Gamma_B={\hbar}/{\tau_B}$; 
$n_4$ is a constant to be fitted in the $\gamma^*\gamma^*\to \jpsi\omega$ coupling, which absorbs the number of the expected $X$ events and corresponding branching fractions; $\rho$ is the phase space factor defined as 
\begin{equation*}
\rho(s,m_1,m_2)=\sqrt{\frac{[s-(m_1-m_2)^2][s-(m_1+m_2)^2]}{s^2}}\ ; \label{phase-space}
\end{equation*}
$m$ ($\Gamma$) denotes the mean mass (width) of a particle~\cite{Zyla:2020zbs}; $a$ (with subscripts) are free parameters.

The fit projections are shown in Fig.~\ref{fig:bubble2}, where 
the fit goodness is gained to be $\chi^2/d.o.f.=56.22/50=1.12$.
Arrayed by signs of phase space factors, a set of Riemann sheets is defined as Table~\ref{tab:bubble2}.
The pole positions in complex $s$ planes are searched for, and also summarized in Table~\ref{tab:bubble2} and sketched in Fig.~\ref{fig:pole_bubble2}.
Only one pole, located on sheet II, is found near the $\Dsp\Dsm$ threshold.
According to PCR~\cite{Morgan:1992ge}, this manifests that
the $X$ structure has molecular $\Dsp\Dsm$ nature.
More exactly, the dynamically molecular picture without the explicit $X$ state is able to describe the current experimental data.
\begin{table}[htbp]   
\begin{ruledtabular}
    \caption{Definition of Riemann sheets for coupled $D_{(s)}\bar{D}_{(s)}$ channels and pole positions. }
    \label{tab:bubble2}
    \begin{tabular}{cccc} 
          Sheet & $\rho_{D\bar{D}}$ & $\rho_{D_s^+D_s^-}$ &Pole position \\
          \hline         
          I   & $+$ &  +  & ...\\
          II  & $-$ &  +  & $3.9207-0.0129i$\\
          III & $-$ & $-$ & ...\\
          IV  & +   & $-$ & ...\\
    \end{tabular}
\end{ruledtabular}
\end{table}

\section{Flatt\'e-like Parameterization}
\label{sec:flatte}

Flatt\'e(-like) formula~\cite{Flatte:1976xu} is a general model with an explicitly-introduced hadron used to parameterize a resonant structure near a hadron-hadron threshold in particle physics, especially in experimental data analyses.
In this subsection, an energy-dependent Flatt\'e-like parameterization for this $X$ state coupling to $\Dp\Dm$, $\Dsp\Dsm$, and $\jpsi\omega$ are used to fit the experimental data and seek pole positions in the complex $s$ planes. PCR~\cite{Morgan:1992ge} and SDFSR~\cite{Weinberg:1962hj,Weinberg:1965zz} are carried out to distinguish whether this $X$ state is a confining state or a molecular $\Dsp\Dsm$ hadron.

\subsection{Parameterization models}

The effective lagrangians of $X$ coupling to the three channels are given by
\begin{equation}
    \begin{aligned}
            \mathcal{L}_{XD^+D^-}&=g_1 XD^+D^-\ , 
            \\
            \mathcal{L}_{XD_s^+ D_s^-}&=g_2XD_s^+ D_s^-\ ,
            \\
            \mathcal{L}_{XJ/\psi \omega}&=g_3X\psi_{\mu} \omega^{\mu}\ .
        \label{flatte lagrangian}
    \end{aligned}
\end{equation}
Then, the corresponding squared amplitudes read
\begin{equation}
    |\mathcal{M}_{B^+ \rightarrow D^+D^- K^+}|^2=\left|i g \frac{i}{s-m_X^2+i m_X\Gamma_X} ig_1\right|^2\ ,\label{bkddfla}
\end{equation}
\begin{equation}
    | \mathcal{M}_{B^+\rightarrow D^+_s D_s^- K^+}|^2=\left|i g \frac{i}{s-m_X^2+i m_X\Gamma_X} ig_2 \right|^2\ ,\label{bkdsfla}
\end{equation}

\begin{equation}
\begin{aligned}
    &| \mathcal{M}_{B^+\rightarrow J/\psi \omega K^+}|^2\\
    &=\left|i g \frac{i}{s-m_X^2+i m_X\Gamma_X} ig_3\right|^2  \\
    & \ \ \times \sum _{pol}\epsilon^*_{\mu}(\Vec{p}_{J/\psi}) \epsilon_{\rho}(\Vec{p}_{J/\psi})\sum_{pol}\epsilon^*_{\nu}(\Vec{p}_{\omega})\epsilon_{\sigma}(\Vec{p}_{\omega})g^{\rho\sigma}g^{\mu\nu}\ ,\label{bkpofla}
\end{aligned}
\end{equation}

\begin{equation}
\begin{aligned}
    &| \mathcal{M}_{\gamma^*\gamma^* \to J/\psi \omega}|^2\\
    &=\left|4ig^\prime (-p_1\cdot p_2 g^{\mu \nu}+p_1^{\nu}p_2^{\mu})\frac{i}{s-m_X^2+i m_X \Gamma_X} i g_3 g^{\rho \sigma}\right|^2 \\
    & \ \ \times \frac{1}{9}\sum_{pol}[\epsilon_{\mu}(\Vec{p}_1)\epsilon^{*}_{\mu^\prime}(\Vec{p}_1)]
     \sum_{pol}[ \epsilon_{\nu}(\Vec{p}_2)  \epsilon^{*}_{\nu^\prime}(\Vec{p}_2) ]\\
     & \ \ \times \sum_{pol}[\epsilon^{*}_{\rho}(\Vec{p}_{J/\psi})\epsilon_{\rho^\prime}(\Vec{p}_{J/\psi})] \sum_{pol}[\epsilon^{*}_{\sigma}(\Vec{p}_{\omega})\epsilon_{\sigma^\prime}(\Vec{p}_{\omega})]\ ,
\end{aligned}
\end{equation}
where $g$ and $g'$ stand for coupling constants of $B^+ \to X K ^+$ and $\gamma^* \gamma^* \to X$, respectively, $\sum_{pol}$ denotes summation of polarization, and 
\begin{equation}
\begin{aligned}
     \Gamma_X&=\frac{1}{16 \pi m_X} \\
    & \ \ \times \left[g_1^2 \rho_1+g_2^2 \rho_2+g_3^2 \rho_3(2+\frac{(s-m_{\psi}^2-m_{\omega}^2)^2}{4 m_{\psi}^2 m_{\omega}^2})\right]\ .
\end{aligned}
\end{equation}

\subsection{Numerical results and pole analysis}

A simultaneous fit to the experimental data~\cite{LHCb-PAPER-2020-025,LHCb-PAPER-2022-018,LHCb-PAPER-2022-019,BaBar:2010wfc,BaBar:2012nxg} is imposed for the mentioned-above decays in this subsection.
The background shapes are parameterized similarly as the $K$-matrix approach.
Then each distribution of the number of events can be expressed as
\begin{equation}
    \begin{aligned}
        &N_{B^+\rightarrow D^+D^- K^+}(s)\\
        &=\frac{\mathcal{R}_{D^+D^-}\mathcal{N}_{X\to D^+D^-}}{\mathcal{B}(B^+ \rightarrow XK^+)\mathcal{B}(X\rightarrow D^+D^-)\Gamma_B}\frac{1}{(2 \pi)^3 32 m_B} \\
        & \ \ \times 2 \sqrt{s} \rho(m_B,m_K,\sqrt{s}) \rho(\sqrt{s},m_D,m_D)   \\
    & \ \ \times 
\left|\frac{g g_1}{D_X(s)}+a_1 e^{i \phi_1}\right|^2\\
                 & \ \ +\left|\frac{g_{\psi(3770)}}{s-m_{\psi(3770)}^2+i m_{\psi(3770)} \Gamma_{\psi(3770)}}\right|^2 \\
    & \ \ +\left|\frac{g_{\chictwo(3930)}}{s-m_{\chictwo(3930)}^2+i m_{\chictwo(3930)} \Gamma_{\chictwo(3930)}}\right|^2\\
    & \ \ +a \sqrt{s} e^{\frac{-(\sqrt{s}-\mu)^2}{2 \sigma^2}}\ ,
    \end{aligned} \label{N_B-KDD}
\end{equation}
\begin{equation}
    \begin{aligned}
       &N_{B^+ \rightarrow D_s^+ D_s^- K^+}(s)\\
       &=\frac{\mathcal{R}_{D_s^+D_s^-}\mathcal{N}_{X\to D_s^+D_s^-}}{\mathcal{B}(B^+\rightarrow XK^+)\mathcal{B}(X\rightarrow D_s^+D_s^-)\Gamma_B} \frac{1}{(2 \pi)^3 32m_B}  \\
        & \ \ \times 2 \sqrt{s} \rho(m_B,m_K,\sqrt{s})\rho(\sqrt{s},m_{D_s},m_{D_s})\\
    & \ \ \times  \left|\frac{g g_2}{D_X(s)}+a_2 e^{i \phi_{21}} \right.\\
    & \ \ \left.+\frac{g_{X_0(4140)} e^{i \phi_{22}}}{s-m_{X_0(4140)}^2+i m_{X_0(4140)}\Gamma_{X_0(4140)}}\right|^2\ , 
    \end{aligned}  \label{N_B-KDs}
\end{equation}
\begin{equation}
    \begin{aligned}
             &N_{B^+\rightarrow J/\psi \omega K^+}(s)\\
             &=\frac{\mathcal{R}_{J/\psi \omega}\mathcal{N}_{X\to\jpsi\omega}}{\mathcal{B}(B^+\rightarrow X K^+)\mathcal{B}(X \rightarrow J/\psi \omega)\Gamma_B}\frac{1}{(2 \pi)^3 32 m_B}  \\
    & \ \ \times 2 \sqrt{s}\rho(m_B,m_K,\sqrt{s})\rho(\sqrt{s},m_{J/\psi},m_{\omega})
              \\
    & \ \ \times \left|\frac{g g_3}{D_X(s)}\right|^2  \left[2+\frac{(s-m_{\psi}^2-m_{\omega}^2)^2}{4 m_{\psi}^2 m_{\omega}^2}\right] \\
    & \ \ +a_{31} \rho(\sqrt{s},m_{J/\psi},m_{\omega})e^{a_{32}(\sqrt{s}-m_{J/\psi}-m_{\omega})}\ ,
    \end{aligned} \label{N_B-KJPO}
\end{equation}
\begin{equation}
\begin{aligned}
    &N_{\gamma^* \gamma^* \rightarrow J/\psi \omega}(s)\\
    &=n_4 \rho(\sqrt{s},m_{J/\psi},m_{\omega})\left|\frac{4 g^\prime g_3}{D_X(s)}\right|^2 \\
    & \ \ \times\left[2+\frac{(s-m_{\psi}^2-m_{\omega}^2)^2}{4 m_{\psi}^2 m_{\omega}^2}\right] \\
    & \ \ \times 
    \frac{1}{9}\left[4q_1^2q_2^2-\frac{(s+q_1^2+q_2^2)^2}{2}+\frac{(s+q_1^2+q_2^2)^4}{16 q_1^2q_2^2}\right]\\
    & \ \ +(a_{41} \sqrt{s}+a_{42})\ ,
\end{aligned}
     \label{N_Gamma}
\end{equation}
where
\begin{equation}
\begin{aligned}
    D_X(s)&=s-m_X^2+i m_X \Gamma_X\ .
\end{aligned}
\end{equation}

\begin{figure*}[htbp]
\centering
\includegraphics[width=0.496\textwidth]{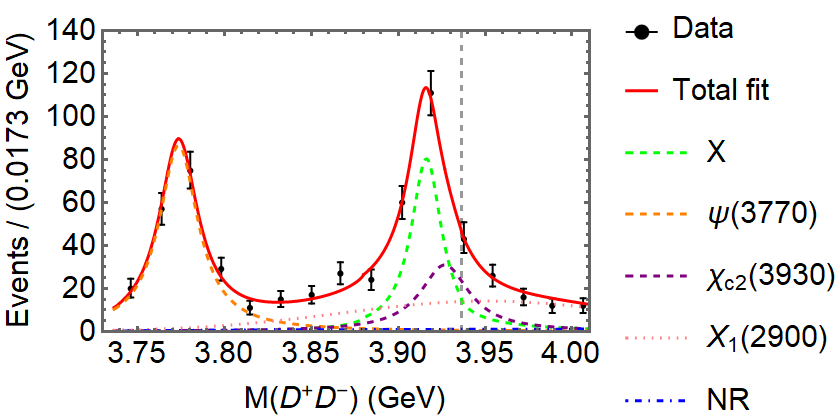}
\includegraphics[width=0.496\textwidth]{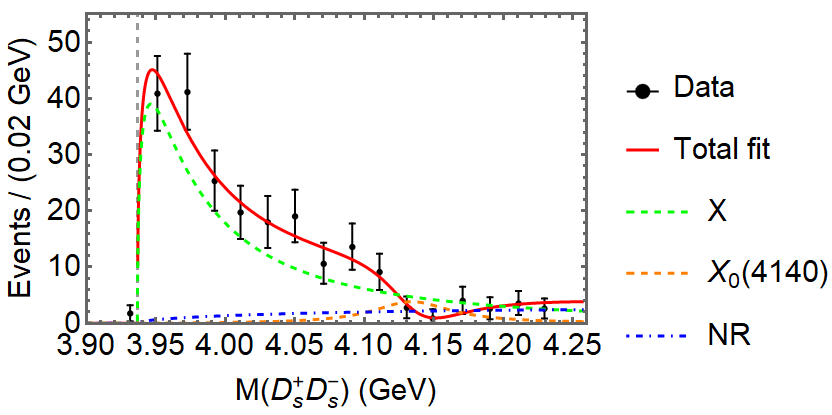} \\
\vspace{0.2cm}
\includegraphics[width=0.45\textwidth]{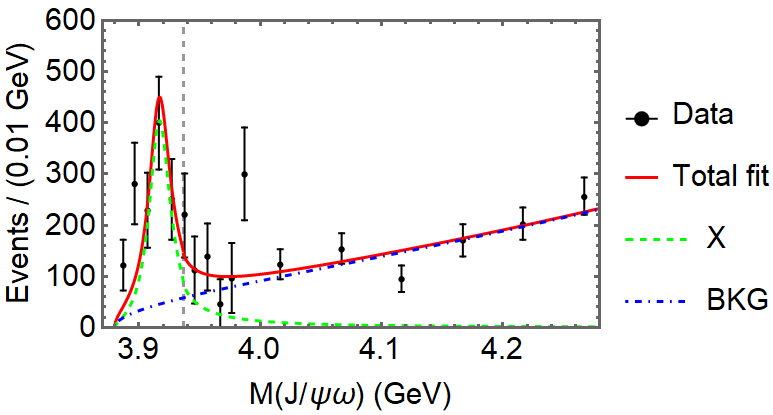}
\hspace{0.7cm}
\includegraphics[width=0.48\textwidth]{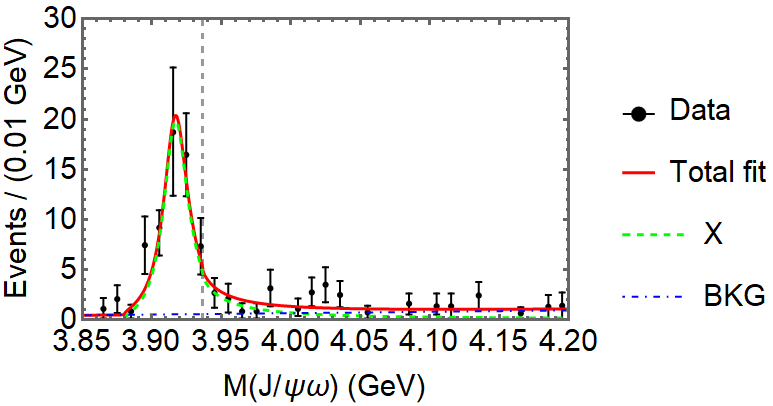}
\hspace{0.2cm}
\caption{(Color online) Fit distributions with Flatt\'e-like parameterizaions for $B^+ \to \Dp\Dm\Kp$ (top left), $B^+ \to D_s^+ D_s^-\Kp$ (top right), $B^+ \to \jpsi \omega\Kp$ (bottom left), and $\gamma^*\gamma^* \to \jpsi \omega$ (bottom right). Here, data are from Ref.~\cite{LHCb-PAPER-2020-025,LHCb-PAPER-2022-018,LHCb-PAPER-2022-019,BaBar:2010wfc,BaBar:2012nxg}, and the vertical dashed lines are located at the $D_s^+ D_s^-$ threshold. }
\label{fig:flatte}
\end{figure*}

\begin{figure*}
    \centering
    \includegraphics[width=17cm]{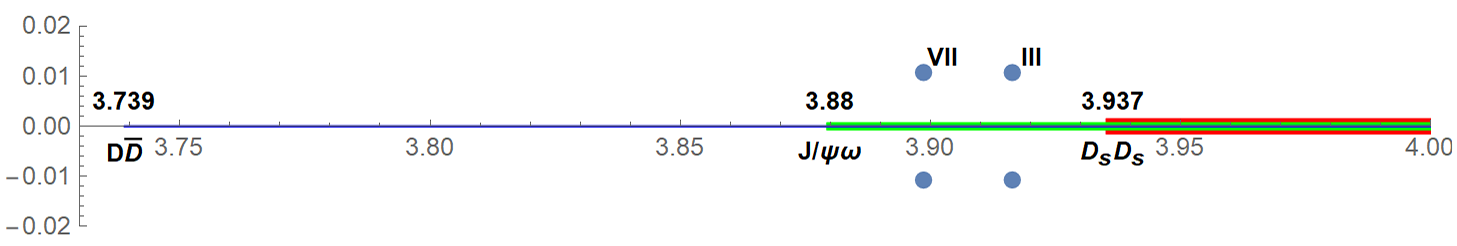}
    \caption{(Color online) Schematic diagram of pole positions for the Flatt\'e-like parameterization.}
    \label{fig:pole_flatte}
\end{figure*}

As shown in Fig.~\ref{fig:flatte}, the fit gives $\chi^2/d.o.f.=52.42/55=0.95$, which is slightly better than the previous $K$-matrix approach.
According to signs of phase space factors, eight Riemann sheets can be generated for the three coupled channels, among which only three sheets have  the largest impact on observables~\cite{PDG2022}, as listed in Table~\ref{tab:flatte} and sketched in Fig.~\ref{fig:pole_flatte}.
The pole positions in complex $s$ planes are searched for and also summarized in Table~\ref{tab:flatte}.
According to PCR~\cite{Morgan:1992ge}, the phenomenon that two poles are found near the $\Dsp\Dsm$ threshold indicates that the $X$ structure gets inclined to attribute with a confining state.
Thus, it can be seen that both the implicit and explicit $X$ interpretations can meet the experimental data well, but the latter is a little better.

\begin{table}[htbp]   
\begin{ruledtabular}
    \caption{Key Riemann sheets for three coupled channels and pole positions. }
    \label{tab:flatte}
    \begin{tabular}{ccccc} 
          Sheet & $\rho_{D\bar{D}}$ & $\rho_{J/\psi \omega}$ &  $\rho_{D_s^+D_s^-}$ & Pole position \\
          \hline         
          II   & $-$ & $+$ & +   & ...\\
          III  & $-$ & $-$ & +   & 3.9163-0.0107$i$\\
          VII  & $-$ & $-$ & $-$ & 3.8986-0.0108$i$\\
    \end{tabular}
\end{ruledtabular}
\end{table}
    
To further gain an insight on the nature of  this near-threshold state,
SDFSR is carried out, which is utilized in an $S$-wave Flatt\'e-like parameterization. From Refs.~\cite{Cao:2020gul,Baru:2003qq,Weinberg:1962hj,Weinberg:1965zz,Kalashnikova:2009gt}, a renormalization constant $\mathcal{Z}$ can be calculated by integrating a spectrum density function with respect to energy, which refers to probability of finding a confining particle in the continuous spectrum: the greater the tendency of $\mathcal{Z}$ to 1, the more the resonant structure is likely to be a confining state; conversely, the closer the value of $\mathcal{Z}$ is to 0, the more the hadron tends to be a hadronic molecule. The fit gives $\mathcal{Z}=0.458$ when the integral interval belongs to $[E_f-\Gamma_X, E_f+\Gamma_X]$; $\mathcal{Z}=0.670$ when $[E_f-2\Gamma_X, E_f+2\Gamma_X]$, where $E_f$ is energy difference between $M_X$ and the $\Dsp\Dsm$ threshold ($m_{th}$), $i.e.$ $E_f=m_X-m_{th}$. The result that the $\mathcal{Z}$ value is slightly less than 0.5 in $[E_f-\Gamma_X, E_f+\Gamma_X]$ but mildly greater than 0.5 in $[E_f-2\Gamma_X, E_f+2\Gamma_X]$, implies that this $X$ state may neither be a pure confining state nor a pure  molecule.
Together with the previous pole analyses, this $X$ resonant structure is more probably a mixture of a confining state and a $\Dsp\Dsm$ hadronic molecule.

\section{Summary and Discussion}
\label{sec:summary}

Based on the assumption that $\chiczero(3930)~(\to D^+ D^-)$, $X(3960)~(\to\Dsp\Dsm)$, and $X(3915)~(\to\jpsi\omega)$ are the same hadron, a combined analysis is performed using both the $K$-matrix approach of $D_{(s)}\bar{D}_{(s)}$ four-point contact interactions and the model of energy-dependent Flatt\'e-like parameterizations.
It is found that both the implicit and explicit $X$ interpretations can meet the experimental data well.
The use of PCR and SDFSR demonstrate that this $X$ hadron is not like a pure $\Dsp\Dsm$ molecule, but might be the mixed nature of a $c\bar{c}$ confining state and $\Dsp\Dsm$ continuum.  
One possible scenario is that the $X$ hadron has a $c\bar{c}$ core strongly renormalized by the $\Dsp\Dsm$ coupling, like the $\chicone(3872)$ as a $c\bar{c}$ resonance with a contribution of the $D^* \bar{D}$ couple-channel effect~\cite{Zhang:2009bv,Meng:2014ota}.

To further analyze the nature of this $X$ state, a number of theoretical predictions for the $^3P$ charmonia are summarized in Table~\ref{tab:3Pj}.
\begin{table*}[htbp]
\centering
\renewcommand\arraystretch{1.2} 
\caption{Summary of $^3P$ charmonia's properties between experimental measurements~\cite{PDG2022}. Here the experimental measurements refer to the mean values from the 2022 Particle Data Group (PDG)~\cite{PDG2022},
and the theoretical predictions in different potential models from  GIM~\cite{Barnes:2005pb}, CPM~\cite{Li:2009ad}, the non-relativistic model~(nRM)~\cite{Barnes:2005pb}, the relativistic perturbative model~(RPM)~\cite{Radford:2007vd}, the relativistic non-perturbative model~(RnPM)~\cite{Radford:2007vd}, the screened potential model~(SPM)~\cite{cc:KTChao},
and the framework of chiral quark model by solving the Schr$\ddot{\rm o}$dinger equation with the Gaussian expansion method~(GEM)~\cite{3Pj:Ping}. The 2022 PDG~\cite{PDG2022} takes $X(3915)$ and $\chiczero(3930)$ as the same particle, denoted as $\chiczero(3915)$. (In units of MeV)}
\label{tab:3Pj}
\begin{tabular*}{\textwidth}{@{\extracolsep{\fill}}cccccccccc}
  \hline
  \multicolumn{2}{c}{State} & EXP & GIM & CPM & nRM  & RPM & RnPM & SPM & GEM \\ \hline
  $1^3P_0$ & $\chiczero(1P)$ & $3414.71\pm0.30$ & 3445 & 3441 & 3424 & 3415.7 & 3415.2 & 3433 & 3430 \\
  $1^3P_1$ & $\chi_{c1}(1P)$ & $3510.67\pm0.05$ & 3510 & 3520 & 3505  & 3508.2 & 3510.6 & 3510 & 3491 \\
  $1^3P_2$ & $\chictwo(2P)$ & $3556.17\pm0.07$ & 3550 & 3565 & 3556  & 3557.7 & 3556.2 & 3554 & 3523 \\
  $2^3P_0$ & \tabincell{c}{$\chiczero(3860)$ \\ $\chiczero(3915)$} & \tabincell{c}{$3862^{+48}_{-35}$ \\ $3921.7\pm1.8$ } & 3916 & 3915 & 3852  & 3843.7 & 3864.3 & 3842 & 3868 \\
  $2^3P_1$ & $\chi_{c1}(3872)$ & $3871.65\pm0.06$ & 3953 & 3875 & 3925 & 3939.7 & 3950.0 & 3901 & 3911 \\
  $2^3P_2$ & $\chictwo(3930)$ & $3922.5\pm1.0$  & 3979 & 3966 & 3972  & 3993.7 & 3992.3 & 3937 & 3935 \\
  $3^3P_0$ &    &   & 4292 &  & 4202 &  &  & 4131 & 4172 \\
  $3^3P_1$ &    &   & 4317 &  & 4271  &  &  & 4178 & 4204 \\
  $3^3P_2$ &    &   & 4337 &  & 4317  &  &  & 4208 & 4222 \\
  \hline
\end{tabular*}
\end{table*}
If this $X$ hadron is indeed a charmonium, it is most likely to be the $\chiczero(2P)$ candidate, which is favored by the  relativistic Godfrey-Isgur model~(GIM)~\cite{Barnes:2005pb}, the couple-channel potential model (CPM)~\cite{Li:2009ad}, and Literature~\cite{Guo:2022zbc}.
However, it is not in agreement with the other theoretical expectations, whose masses are predicted in the range of 3842–3868 MeV~\cite{Barnes:2005pb,Radford:2007vd,cc:KTChao,3Pj:Ping}.
Another phenomenon is that two candidates can be treated as the $\chiczero(2P)$ charmonium: $\chiczero(3860)$~\cite{Belle:2017egg} discovered in the $D\bar{D}$ decays via $e^+e^-\to\jpsi D \bar{D}$ and the $X$ state discussed in this work.
Yet Ref.~\cite{Deineka:2021aeu} argued that  the $\chi_{c0}(3860)$ peak is due to a bound state around 3695 MeV. It needs to be confirmed in the forthcoming experimental measurements.
\begin{table}[htbp]   
\begin{ruledtabular}
    \caption{Pole positions without the $\jpsi\omega$ channel. }
    \label{tab:non3915}
    \begin{tabular}{lcccc} 
        Case & Sheet & $\rho_{D\bar{D}}$ & $\rho_{D_s^+D_s^-}$ &Pole position \\
          \hline         
        $K$-matrix  & II  & $-$ &  +  & $3.9279-0.0079i$\\ \hline
         \multirow{2}*{Flatt\'e-like} & II & $-$ & $+$ & $3.9303-0.0041i$ \\
          & III  & $-$ & $-$ & $3.8702-0.0109i$ \\
    \end{tabular}
\end{ruledtabular}
\end{table}
\begin{table}[htbp]   
\begin{ruledtabular}
    \caption{Pole positions without the $\Dsp \Dsm$ channel. }
    \label{tab:non3960}
    \begin{tabular}{lcccc} 
        Case & Sheet & $\rho_{D\bar{D}}$ & $\rho_{\jpsi\omega}$ &Pole position \\
          \hline         
        $K$-matrix  & II  & $-$ &  +  & $3.9189-0.0115i$\\ \hline
         \multirow{2}*{Flatt\'e-like} & II & $-$ & $+$ & $3.9164-0.0046i$ \\
          & III  & $-$ & $-$ & $3.9157-0.0101i$ \\
    \end{tabular}
\end{ruledtabular}
\end{table}
Whatever, more accurate studies based on potential models, as well as other methods, are needed to shed light on the nature of the $X$ state.
For example, Ref.~\cite{Xie:2022lyw} estimated the branching fraction of $B^+ \to X(3960) K^+$ to be $(2.9-13.3)\times10^{-4}$ if assuming $X(3960)$ as a $\Dsp\Dsm$ bound state, which can be helpful in the future experiments to test if the $X(3960)$ hadron is a bound state.

Due to limited data statistics, however, we cannot draw a solid conclusion in this work.
More experimental data are expected
to further clarify the nature of $\chiczero(3930)$/$X(3960)$/$X(3915)$, for instance, the $\gamma\gamma\to D_{(s)}\bar{D}_{(s)}$ reactions, the $e^+e^-\to\psi D_{(s)}\bar{D}_{(s)}$ productions, the amplitude analysis for the $X(3915)\to\jpsi\omega$ chain, and the ratio of $\Gamma(X\to D_{(s)}\bar{D}_{(s)})/\Gamma(X\to\jpsi\omega)$. 
Without doubt, other decay modes are also valuable to elucidate the nature of the $\Dsp\Dsm$ near-threshold structure, such as $X\to  \eta^{(\prime)}\eta_c,  \pi\pi\chi_{c0,2}, \gamma \jpsi, \gamma \psi(3686), $ $\gamma\psi(3770), \gamma D^{(*)} \bar{D}, \gamma \Dsp\Dsm$, etc.

Nevertheless, it is noteworthy that the $0^{++}$ assignment for the $X(3915)(\to\jpsi\omega)$ state is not completely determined by experiments. Several works take the $X(3915)(\to\jpsi\omega)$ as the $2^{++}$ charmonium $\chictwo(3930)$~\cite{Ji:2022vdj,Zhou:2015uva}, but the $\chiczero(3930)~(\to D^+ D^-)$ and $X(3960)~(\to\Dsp\Dsm)$ are the same $0^{++}$ molecular hadron~\cite{Ji:2022vdj}. In view of this assumption,  fits without the $\jpsi\omega$ channel are also tested, where the numerical results are summarized in Table~\ref{tab:non3915}. These pole positions are roughly consistent  with the nominal results though the elementariness of $\chiczero(3930)/X(3960)$ is less favored here.
In addition, Refs.~\cite{Ji:2022vdj,Badalian:2023qyi} regards the $X(3960)$ as a different state from $\chiczero(3930)~(\to D^+ D^-)$/$X(3915)(\to\jpsi\omega)$, so that fits without the $\Dsp\Dsm$ decay are used to check. As listed in Table~\ref{tab:non3960}, the numerical values are compatible with the nominal ones, which shed light on the mixed nature of $X(3915)/\chiczero(3930)$. 
That is, it does not shake the conclusion of this article in case that the three decays are not from the same hadron.


\begin{acknowledgments}
This work is supported in part by National Nature Science Foundations
of China under Contract Number 11975028, 10925522 and 11875071.
\end{acknowledgments}

\appendix

\section{Relevant fitted parameters}

The parameters associated with pole positions are summarized in Tables~\ref{tab:par-kmatrix} and \ref{tab:par-flatte}.

\begin{table}[htbp]   
\begin{ruledtabular}
    \caption{Parameters associated with pole positions in the $K$-matrix method, which are $g_{11},~g_{12},~g_{22}$ in Eq.~\ref{KI=0} and subtraction constant $a(\mu)$ of $B_0$ function in Eq.~\ref{B0 function}, 
     where $g_{11}=g(1+\delta_1),~g_{22}=g(1+\delta_2)$, and $g_{12}=2g$.}
    \label{tab:par-kmatrix}
    \begin{tabular}{ll} 
Parameter & Value  \\
\hline
$\delta_1$      & $-0.0450\pm0.0653$            \\
$\delta_2$      & $2.4954\pm1.4267$             \\
$g$             & $1.6543\pm0.6575$             \\
$a(\mu)$        & $8.0516\pm2.4454$             \\
    \end{tabular}
\end{ruledtabular}
\end{table}

\begin{table}[htbp]   
\begin{ruledtabular}
    \caption{Parameters associated with pole positions in Flatt\'e-like parameterizations.}
    \label{tab:par-flatte}
    \begin{tabular}{ll} 
Parameter & Value  \\
\hline
$m_X~(\text{GeV})$           & $3.9108\pm0.0104$              \\
$g_1~(\text{GeV})$           & $1.2287\pm1.9256$              \\
$g_2~(\text{GeV})$           & $5.2372\pm4.0788$              \\
$g_3~(\text{GeV})$           & $3.6111\pm0.6263$              \\
    \end{tabular}
\end{ruledtabular}
\end{table}

\nocite{*}

\bibliography{main}

\begin{thebibliography}{45}%
\makeatletter
\providecommand \@ifxundefined [1]{%
 \@ifx{#1\undefined}
}%
\providecommand \@ifnum [1]{%
 \ifnum #1\expandafter \@firstoftwo
 \else \expandafter \@secondoftwo
 \fi
}%
\providecommand \@ifx [1]{%
 \ifx #1\expandafter \@firstoftwo
 \else \expandafter \@secondoftwo
 \fi
}%
\providecommand \natexlab [1]{#1}%
\providecommand \enquote  [1]{``#1''}%
\providecommand \bibnamefont  [1]{#1}%
\providecommand \bibfnamefont [1]{#1}%
\providecommand \citenamefont [1]{#1}%
\providecommand \href@noop [0]{\@secondoftwo}%
\providecommand \href [0]{\begingroup \@sanitize@url \@href}%
\providecommand \@href[1]{\@@startlink{#1}\@@href}%
\providecommand \@@href[1]{\endgroup#1\@@endlink}%
\providecommand \@sanitize@url [0]{\catcode `\\12\catcode `\$12\catcode
  `\&12\catcode `\#12\catcode `\^12\catcode `\_12\catcode `\%12\relax}%
\providecommand \@@startlink[1]{}%
\providecommand \@@endlink[0]{}%
\providecommand \url  [0]{\begingroup\@sanitize@url \@url }%
\providecommand \@url [1]{\endgroup\@href {#1}{\urlprefix }}%
\providecommand \urlprefix  [0]{URL }%
\providecommand \Eprint [0]{\href }%
\providecommand \doibase [0]{https://doi.org/}%
\providecommand \selectlanguage [0]{\@gobble}%
\providecommand \bibinfo  [0]{\@secondoftwo}%
\providecommand \bibfield  [0]{\@secondoftwo}%
\providecommand \translation [1]{[#1]}%
\providecommand \BibitemOpen [0]{}%
\providecommand \bibitemStop [0]{}%
\providecommand \bibitemNoStop [0]{.\EOS\space}%
\providecommand \EOS [0]{\spacefactor3000\relax}%
\providecommand \BibitemShut  [1]{\csname bibitem#1\endcsname}%
\let\auto@bib@innerbib\@empty
\bibitem [{\citenamefont {Aaij}\ \emph
  {et~al.}(2022{\natexlab{a}})\citenamefont {Aaij} \emph
  {et~al.}}]{LHCb-PAPER-2022-019}%
  \BibitemOpen
  \bibfield  {author} {\bibinfo {author} {\bibfnamefont {R.}~\bibnamefont
  {Aaij}} \emph {et~al.} (\bibinfo {collaboration} {LHCb collaboration}),\
  }\bibfield  {title} {\bibinfo {title} {{First observation of the $B^+ \to
  D^+_s D^+_s K^+$ decay}},\ }\href@noop {} {\bibfield  {journal} {\bibinfo
  {journal} {Phys. Rev.}\ }\textbf {\bibinfo {volume} {Dxxx}} (\bibinfo {year}
  {2022}{\natexlab{a}})},\ \Eprint {https://arxiv.org/abs/2211.05034}
  {arXiv:2211.05034 [hep-ex]} \BibitemShut {NoStop}%
\bibitem [{\citenamefont {Aaij}\ \emph
  {et~al.}(2022{\natexlab{b}})\citenamefont {Aaij} \emph
  {et~al.}}]{LHCb-PAPER-2022-018}%
  \BibitemOpen
  \bibfield  {author} {\bibinfo {author} {\bibfnamefont {R.}~\bibnamefont
  {Aaij}} \emph {et~al.} (\bibinfo {collaboration} {LHCb collaboration}),\
  }\bibfield  {title} {\bibinfo {title} {{Observation of a resonant structure
  near the $D^{+}_{s} D^{-}_{s}$ threshold}},\ }\href@noop {} {\bibfield
  {journal} {\bibinfo  {journal} {Phys. Rev. Lett.}\ }\textbf {\bibinfo
  {volume} {xxx}} (\bibinfo {year} {2022}{\natexlab{b}})},\ \Eprint
  {https://arxiv.org/abs/2210.15153} {arXiv:2210.15153 [hep-ex]} \BibitemShut
  {NoStop}%
\bibitem [{\citenamefont {Aaij}\ \emph {et~al.}(2020)\citenamefont {Aaij} \emph
  {et~al.}}]{LHCb-PAPER-2020-025}%
  \BibitemOpen
  \bibfield  {author} {\bibinfo {author} {\bibfnamefont {R.}~\bibnamefont
  {Aaij}} \emph {et~al.} (\bibinfo {collaboration} {LHCb collaboration}),\
  }\bibfield  {title} {\bibinfo {title} {{Amplitude analysis of the $B^+ \to
  D^+ D^- K^+$ decay}},\ }\href {https://doi.org/10.1103/PhysRevD.102.112003}
  {\bibfield  {journal} {\bibinfo  {journal} {Phys. Rev.}\ }\textbf {\bibinfo
  {volume} {D102}},\ \bibinfo {pages} {112003} (\bibinfo {year} {2020})},\
  \Eprint {https://arxiv.org/abs/2009.00026} {arXiv:2009.00026 [hep-ex]}
  \BibitemShut {NoStop}%
\bibitem [{\citenamefont {Bayar}\ \emph {et~al.}(2023)\citenamefont {Bayar},
  \citenamefont {Feijoo},\ and\ \citenamefont {Oset}}]{Bayar:2022dqa}%
  \BibitemOpen
  \bibfield  {author} {\bibinfo {author} {\bibfnamefont {M.}~\bibnamefont
  {Bayar}}, \bibinfo {author} {\bibfnamefont {A.}~\bibnamefont {Feijoo}},\ and\
  \bibinfo {author} {\bibfnamefont {E.}~\bibnamefont {Oset}},\ }\bibfield
  {title} {\bibinfo {title} {{X(3960) seen in Ds+Ds- as the X(3930) state seen
  in D+D-}},\ }\href {https://doi.org/10.1103/PhysRevD.107.034007} {\bibfield
  {journal} {\bibinfo  {journal} {Phys. Rev. D}\ }\textbf {\bibinfo {volume}
  {107}},\ \bibinfo {pages} {034007} (\bibinfo {year} {2023})},\ \Eprint
  {https://arxiv.org/abs/2207.08490} {arXiv:2207.08490 [hep-ph]} \BibitemShut
  {NoStop}%
\bibitem [{\citenamefont {Ji}\ \emph {et~al.}(2022{\natexlab{a}})\citenamefont
  {Ji}, \citenamefont {Dong}, \citenamefont {Albaladejo}, \citenamefont {Du},
  \citenamefont {Guo}, \citenamefont {Nieves},\ and\ \citenamefont
  {Zou}}]{Ji:2022vdj}%
  \BibitemOpen
  \bibfield  {author} {\bibinfo {author} {\bibfnamefont {T.}~\bibnamefont
  {Ji}}, \bibinfo {author} {\bibfnamefont {X.-K.}\ \bibnamefont {Dong}},
  \bibinfo {author} {\bibfnamefont {M.}~\bibnamefont {Albaladejo}}, \bibinfo
  {author} {\bibfnamefont {M.-L.}\ \bibnamefont {Du}}, \bibinfo {author}
  {\bibfnamefont {F.-K.}\ \bibnamefont {Guo}}, \bibinfo {author} {\bibfnamefont
  {J.}~\bibnamefont {Nieves}},\ and\ \bibinfo {author} {\bibfnamefont {B.-S.}\
  \bibnamefont {Zou}},\ }\bibfield  {title} {\bibinfo {title} {{Understanding
  the $0^{++}$ and $2^{++}$ charmonium(-like) states near 3.9 GeV}},\
  }\href@noop {} {\bibfield  {journal} {\bibinfo  {journal} {xxx}\ } (\bibinfo
  {year} {2022}{\natexlab{a}})},\ \Eprint {https://arxiv.org/abs/2212.00631}
  {arXiv:2212.00631 [hep-ph]} \BibitemShut {NoStop}%
\bibitem [{\citenamefont {Xin}\ \emph {et~al.}(2022)\citenamefont {Xin},
  \citenamefont {Wang},\ and\ \citenamefont {Yang}}]{Xin:2022bzt}%
  \BibitemOpen
  \bibfield  {author} {\bibinfo {author} {\bibfnamefont {Q.}~\bibnamefont
  {Xin}}, \bibinfo {author} {\bibfnamefont {Z.-G.}\ \bibnamefont {Wang}},\ and\
  \bibinfo {author} {\bibfnamefont {X.-S.}\ \bibnamefont {Yang}},\ }\bibfield
  {title} {\bibinfo {title} {{Analysis of the $X(3960)$ and related tetraquark
  molecular states via the QCD sum rules}},\ }\href@noop {} {\bibfield
  {journal} {\bibinfo  {journal} {xxx}\ } (\bibinfo {year} {2022})},\ \Eprint
  {https://arxiv.org/abs/2207.09910} {arXiv:2207.09910 [hep-ph]} \BibitemShut
  {NoStop}%
\bibitem [{\citenamefont {Chen}\ and\ \citenamefont
  {Huang}(2022)}]{Chen:2022dad}%
  \BibitemOpen
  \bibfield  {author} {\bibinfo {author} {\bibfnamefont {R.}~\bibnamefont
  {Chen}}\ and\ \bibinfo {author} {\bibfnamefont {Q.}~\bibnamefont {Huang}},\
  }\bibfield  {title} {\bibinfo {title} {{Charmoniumlike resonant explanation
  on the newly observed $X(3960)$}},\ }\href@noop {} {\bibfield  {journal}
  {\bibinfo  {journal} {xxx}\ } (\bibinfo {year} {2022})},\ \Eprint
  {https://arxiv.org/abs/2209.05180} {arXiv:2209.05180 [hep-ph]} \BibitemShut
  {NoStop}%
\bibitem [{\citenamefont {Mutuk}(2022)}]{Mutuk:2022ckn}%
  \BibitemOpen
  \bibfield  {author} {\bibinfo {author} {\bibfnamefont {H.}~\bibnamefont
  {Mutuk}},\ }\bibfield  {title} {\bibinfo {title} {{Molecular interpretation
  of X(3960) as $D_s^+ D_s^-$ state}},\ }\href
  {https://doi.org/10.1140/epjc/s10052-022-11120-3} {\bibfield  {journal}
  {\bibinfo  {journal} {Eur. Phys. J. C}\ }\textbf {\bibinfo {volume} {82}},\
  \bibinfo {pages} {1142} (\bibinfo {year} {2022})},\ \Eprint
  {https://arxiv.org/abs/2211.14836} {arXiv:2211.14836 [hep-ph]} \BibitemShut
  {NoStop}%
\bibitem [{\citenamefont {Agaev}\ \emph {et~al.}(2022)\citenamefont {Agaev},
  \citenamefont {Azizi},\ and\ \citenamefont {Sundu}}]{Agaev:2022pis}%
  \BibitemOpen
  \bibfield  {author} {\bibinfo {author} {\bibfnamefont {S.~S.}\ \bibnamefont
  {Agaev}}, \bibinfo {author} {\bibfnamefont {K.}~\bibnamefont {Azizi}},\ and\
  \bibinfo {author} {\bibfnamefont {H.}~\bibnamefont {Sundu}},\ }\bibfield
  {title} {\bibinfo {title} {{Resonance $X(3960)$ as a hidden charm-strange
  scalar tetraquark}},\ }\href@noop {} {\bibfield  {journal} {\bibinfo
  {journal} {xxx}\ } (\bibinfo {year} {2022})},\ \Eprint
  {https://arxiv.org/abs/2211.14129} {arXiv:2211.14129 [hep-ph]} \BibitemShut
  {NoStop}%
\bibitem [{\citenamefont {Guo}\ \emph {et~al.}(2022{\natexlab{a}})\citenamefont
  {Guo}, \citenamefont {Li}, \citenamefont {Zhao},\ and\ \citenamefont
  {He}}]{Guo:2022crh}%
  \BibitemOpen
  \bibfield  {author} {\bibinfo {author} {\bibfnamefont {T.}~\bibnamefont
  {Guo}}, \bibinfo {author} {\bibfnamefont {J.}~\bibnamefont {Li}}, \bibinfo
  {author} {\bibfnamefont {J.}~\bibnamefont {Zhao}},\ and\ \bibinfo {author}
  {\bibfnamefont {L.}~\bibnamefont {He}},\ }\bibfield  {title} {\bibinfo
  {title} {{Investigation of the tetraquark states $Qq\bar{Q} \bar{q}$ in the
  improved chromomagnetic interaction model}},\ }\href@noop {} {\bibfield
  {journal} {\bibinfo  {journal} {xxx}\ } (\bibinfo {year}
  {2022}{\natexlab{a}})},\ \Eprint {https://arxiv.org/abs/2211.10834}
  {arXiv:2211.10834 [hep-ph]} \BibitemShut {NoStop}%
\bibitem [{\citenamefont {Badalian}\ and\ \citenamefont
  {Simonov}(2023)}]{Badalian:2023qyi}%
  \BibitemOpen
  \bibfield  {author} {\bibinfo {author} {\bibfnamefont {A.~M.}\ \bibnamefont
  {Badalian}}\ and\ \bibinfo {author} {\bibfnamefont {Y.~A.}\ \bibnamefont
  {Simonov}},\ }\bibfield  {title} {\bibinfo {title} {{The scalar exotic
  resonances X(3915), X(3960), X(4140)}},\ }\href@noop {} {\bibfield  {journal}
  {\bibinfo  {journal} {xxx}\ } (\bibinfo {year} {2023})},\ \Eprint
  {https://arxiv.org/abs/2301.13597} {arXiv:2301.13597 [hep-ph]} \BibitemShut
  {NoStop}%
\bibitem [{\citenamefont {Workman}\ \emph {et~al.}(2022)\citenamefont {Workman}
  \emph {et~al.}}]{PDG2022}%
  \BibitemOpen
  \bibfield  {author} {\bibinfo {author} {\bibfnamefont {R.~L.}\ \bibnamefont
  {Workman}} \emph {et~al.} (\bibinfo {collaboration} {Particle Data Group}),\
  }\bibfield  {title} {\bibinfo {title} {{Review of particle physics}},\ }\href
  {https://doi.org/10.1093/ptep/ptac097} {\bibfield  {journal} {\bibinfo
  {journal} {Prog. Theor. Exp. Phys.}\ }\textbf {\bibinfo {volume} {2022}},\
  \bibinfo {pages} {083C01} (\bibinfo {year} {2022})}\BibitemShut {NoStop}%
\bibitem [{\citenamefont {Morgan}(1992)}]{Morgan:1992ge}%
  \BibitemOpen
  \bibfield  {author} {\bibinfo {author} {\bibfnamefont {D.}~\bibnamefont
  {Morgan}},\ }\bibfield  {title} {\bibinfo {title} {{Pole counting and
  resonance classification}},\ }\href
  {https://doi.org/10.1016/0375-9474(92)90550-4} {\bibfield  {journal}
  {\bibinfo  {journal} {Nucl. Phys. A}\ }\textbf {\bibinfo {volume} {543}},\
  \bibinfo {pages} {632} (\bibinfo {year} {1992})}\BibitemShut {NoStop}%
\bibitem [{\citenamefont {Zhang}\ \emph {et~al.}(2009)\citenamefont {Zhang},
  \citenamefont {Meng},\ and\ \citenamefont {Zheng}}]{Zhang:2009bv}%
  \BibitemOpen
  \bibfield  {author} {\bibinfo {author} {\bibfnamefont {O.}~\bibnamefont
  {Zhang}}, \bibinfo {author} {\bibfnamefont {C.}~\bibnamefont {Meng}},\ and\
  \bibinfo {author} {\bibfnamefont {H.~Q.}\ \bibnamefont {Zheng}},\ }\bibfield
  {title} {\bibinfo {title} {{Ambiversion of X(3872)}},\ }\href
  {https://doi.org/10.1016/j.physletb.2009.09.033} {\bibfield  {journal}
  {\bibinfo  {journal} {Phys. Lett. B}\ }\textbf {\bibinfo {volume} {680}},\
  \bibinfo {pages} {453} (\bibinfo {year} {2009})},\ \Eprint
  {https://arxiv.org/abs/0901.1553} {arXiv:0901.1553 [hep-ph]} \BibitemShut
  {NoStop}%
\bibitem [{\citenamefont {Dai}\ \emph {et~al.}(2015)\citenamefont {Dai},
  \citenamefont {Shi}, \citenamefont {Tang},\ and\ \citenamefont
  {Zheng}}]{Dai:2012pb}%
  \BibitemOpen
  \bibfield  {author} {\bibinfo {author} {\bibfnamefont {L.~Y.}\ \bibnamefont
  {Dai}}, \bibinfo {author} {\bibfnamefont {M.}~\bibnamefont {Shi}}, \bibinfo
  {author} {\bibfnamefont {G.-Y.}\ \bibnamefont {Tang}},\ and\ \bibinfo
  {author} {\bibfnamefont {H.~Q.}\ \bibnamefont {Zheng}},\ }\bibfield  {title}
  {\bibinfo {title} {{Nature of X(4260)}},\ }\href
  {https://doi.org/10.1103/PhysRevD.92.014020} {\bibfield  {journal} {\bibinfo
  {journal} {Phys. Rev. D}\ }\textbf {\bibinfo {volume} {92}},\ \bibinfo
  {pages} {014020} (\bibinfo {year} {2015})},\ \Eprint
  {https://arxiv.org/abs/1206.6911} {arXiv:1206.6911 [hep-ph]} \BibitemShut
  {NoStop}%
\bibitem [{\citenamefont {Cao}\ \emph {et~al.}(2019)\citenamefont {Cao},
  \citenamefont {Qi}, \citenamefont {Wang},\ and\ \citenamefont
  {Zheng}}]{Cao:2019wwt}%
  \BibitemOpen
  \bibfield  {author} {\bibinfo {author} {\bibfnamefont {Q.-F.}\ \bibnamefont
  {Cao}}, \bibinfo {author} {\bibfnamefont {H.-R.}\ \bibnamefont {Qi}},
  \bibinfo {author} {\bibfnamefont {Y.-F.}\ \bibnamefont {Wang}},\ and\
  \bibinfo {author} {\bibfnamefont {H.-Q.}\ \bibnamefont {Zheng}},\ }\bibfield
  {title} {\bibinfo {title} {{Discussions on the line-shape of the $X$(4660)
  resonance}},\ }\href {https://doi.org/10.1103/PhysRevD.100.054040} {\bibfield
   {journal} {\bibinfo  {journal} {Phys. Rev. D}\ }\textbf {\bibinfo {volume}
  {100}},\ \bibinfo {pages} {054040} (\bibinfo {year} {2019})},\ \Eprint
  {https://arxiv.org/abs/1906.00356} {arXiv:1906.00356 [hep-ph]} \BibitemShut
  {NoStop}%
\bibitem [{\citenamefont {Chen}\ \emph {et~al.}(2021)\citenamefont {Chen},
  \citenamefont {Qi},\ and\ \citenamefont {Zheng}}]{Chen:2021tad}%
  \BibitemOpen
  \bibfield  {author} {\bibinfo {author} {\bibfnamefont {H.}~\bibnamefont
  {Chen}}, \bibinfo {author} {\bibfnamefont {H.-R.}\ \bibnamefont {Qi}},\ and\
  \bibinfo {author} {\bibfnamefont {H.-Q.}\ \bibnamefont {Zheng}},\ }\bibfield
  {title} {\bibinfo {title} {{$X_1(2900)$ as a $\bar{D}_1 K$ molecule}},\
  }\href {https://doi.org/10.1140/epjc/s10052-021-09603-w} {\bibfield
  {journal} {\bibinfo  {journal} {Eur. Phys. J. C}\ }\textbf {\bibinfo {volume}
  {81}},\ \bibinfo {pages} {812} (\bibinfo {year} {2021})},\ \Eprint
  {https://arxiv.org/abs/2108.02387} {arXiv:2108.02387 [hep-ph]} \BibitemShut
  {NoStop}%
\bibitem [{\citenamefont {Gong}\ \emph {et~al.}(2016)\citenamefont {Gong},
  \citenamefont {Guo}, \citenamefont {Meng}, \citenamefont {Tang},
  \citenamefont {Wang},\ and\ \citenamefont {Zheng}}]{Gong:2016hlt}%
  \BibitemOpen
  \bibfield  {author} {\bibinfo {author} {\bibfnamefont {Q.-R.}\ \bibnamefont
  {Gong}}, \bibinfo {author} {\bibfnamefont {Z.-H.}\ \bibnamefont {Guo}},
  \bibinfo {author} {\bibfnamefont {C.}~\bibnamefont {Meng}}, \bibinfo {author}
  {\bibfnamefont {G.-Y.}\ \bibnamefont {Tang}}, \bibinfo {author}
  {\bibfnamefont {Y.-F.}\ \bibnamefont {Wang}},\ and\ \bibinfo {author}
  {\bibfnamefont {H.-Q.}\ \bibnamefont {Zheng}},\ }\bibfield  {title} {\bibinfo
  {title} {{$Z_c(3900)$ as a $D\bar{D}^*$ molecule from the pole counting
  rule}},\ }\href {https://doi.org/10.1103/PhysRevD.94.114019} {\bibfield
  {journal} {\bibinfo  {journal} {Phys. Rev. D}\ }\textbf {\bibinfo {volume}
  {94}},\ \bibinfo {pages} {114019} (\bibinfo {year} {2016})},\ \Eprint
  {https://arxiv.org/abs/1604.08836} {arXiv:1604.08836 [hep-ph]} \BibitemShut
  {NoStop}%
\bibitem [{\citenamefont {Cao}\ \emph {et~al.}(2021{\natexlab{a}})\citenamefont
  {Cao}, \citenamefont {Chen}, \citenamefont {Qi},\ and\ \citenamefont
  {Zheng}}]{Cao:2020gul}%
  \BibitemOpen
  \bibfield  {author} {\bibinfo {author} {\bibfnamefont {Q.-F.}\ \bibnamefont
  {Cao}}, \bibinfo {author} {\bibfnamefont {H.}~\bibnamefont {Chen}}, \bibinfo
  {author} {\bibfnamefont {H.-R.}\ \bibnamefont {Qi}},\ and\ \bibinfo {author}
  {\bibfnamefont {H.-Q.}\ \bibnamefont {Zheng}},\ }\bibfield  {title} {\bibinfo
  {title} {{Some remarks on $X(6900)$}},\ }\href
  {https://doi.org/10.1088/1674-1137/ac0ee5} {\bibfield  {journal} {\bibinfo
  {journal} {Chin. Phys. C}\ }\textbf {\bibinfo {volume} {45}},\ \bibinfo
  {pages} {103102} (\bibinfo {year} {2021}{\natexlab{a}})},\ \Eprint
  {https://arxiv.org/abs/2011.04347} {arXiv:2011.04347 [hep-ph]} \BibitemShut
  {NoStop}%
\bibitem [{\citenamefont {Baru}\ \emph {et~al.}(2004)\citenamefont {Baru},
  \citenamefont {Haidenbauer}, \citenamefont {Hanhart}, \citenamefont
  {Kalashnikova},\ and\ \citenamefont {Kudryavtsev}}]{Baru:2003qq}%
  \BibitemOpen
  \bibfield  {author} {\bibinfo {author} {\bibfnamefont {V.}~\bibnamefont
  {Baru}}, \bibinfo {author} {\bibfnamefont {J.}~\bibnamefont {Haidenbauer}},
  \bibinfo {author} {\bibfnamefont {C.}~\bibnamefont {Hanhart}}, \bibinfo
  {author} {\bibfnamefont {Y.}~\bibnamefont {Kalashnikova}},\ and\ \bibinfo
  {author} {\bibfnamefont {A.~E.}\ \bibnamefont {Kudryavtsev}},\ }\bibfield
  {title} {\bibinfo {title} {{Evidence that the a(0)(980) and f(0)(980) are not
  elementary particles}},\ }\href
  {https://doi.org/10.1016/j.physletb.2004.01.088} {\bibfield  {journal}
  {\bibinfo  {journal} {Phys. Lett. B}\ }\textbf {\bibinfo {volume} {586}},\
  \bibinfo {pages} {53} (\bibinfo {year} {2004})},\ \Eprint
  {https://arxiv.org/abs/hep-ph/0308129} {arXiv:hep-ph/0308129} \BibitemShut
  {NoStop}%
\bibitem [{\citenamefont {Weinberg}(1963)}]{Weinberg:1962hj}%
  \BibitemOpen
  \bibfield  {author} {\bibinfo {author} {\bibfnamefont {S.}~\bibnamefont
  {Weinberg}},\ }\bibfield  {title} {\bibinfo {title} {{Elementary particle
  theory of composite particles}},\ }\href
  {https://doi.org/10.1103/PhysRev.130.776} {\bibfield  {journal} {\bibinfo
  {journal} {Phys. Rev.}\ }\textbf {\bibinfo {volume} {130}},\ \bibinfo {pages}
  {776} (\bibinfo {year} {1963})}\BibitemShut {NoStop}%
\bibitem [{\citenamefont {Weinberg}(1965)}]{Weinberg:1965zz}%
  \BibitemOpen
  \bibfield  {author} {\bibinfo {author} {\bibfnamefont {S.}~\bibnamefont
  {Weinberg}},\ }\bibfield  {title} {\bibinfo {title} {{Evidence That the
  Deuteron Is Not an Elementary Particle}},\ }\href
  {https://doi.org/10.1103/PhysRev.137.B672} {\bibfield  {journal} {\bibinfo
  {journal} {Phys. Rev.}\ }\textbf {\bibinfo {volume} {137}},\ \bibinfo {pages}
  {B672} (\bibinfo {year} {1965})}\BibitemShut {NoStop}%
\bibitem [{\citenamefont {Kalashnikova}\ and\ \citenamefont
  {Nefediev}(2009)}]{Kalashnikova:2009gt}%
  \BibitemOpen
  \bibfield  {author} {\bibinfo {author} {\bibfnamefont {Y.~S.}\ \bibnamefont
  {Kalashnikova}}\ and\ \bibinfo {author} {\bibfnamefont {A.~V.}\ \bibnamefont
  {Nefediev}},\ }\bibfield  {title} {\bibinfo {title} {{Nature of X(3872) from
  data}},\ }\href {https://doi.org/10.1103/PhysRevD.80.074004} {\bibfield
  {journal} {\bibinfo  {journal} {Phys. Rev. D}\ }\textbf {\bibinfo {volume}
  {80}},\ \bibinfo {pages} {074004} (\bibinfo {year} {2009})},\ \Eprint
  {https://arxiv.org/abs/0907.4901} {arXiv:0907.4901 [hep-ph]} \BibitemShut
  {NoStop}%
\bibitem [{\citenamefont {Du}(2017)}]{Du:phd}%
  \BibitemOpen
  \bibfield  {author} {\bibinfo {author} {\bibfnamefont {M.~L.}\ \bibnamefont
  {Du}},\ }\emph {\bibinfo {title} {{Topics in chiral perturbation theory for
  charmed mesons}}},\ \href@noop {} {Ph.D. thesis},\ \bibinfo  {school} {Bonn
  U.} (\bibinfo {year} {2017})\BibitemShut {NoStop}%
\bibitem [{\citenamefont {Oller}\ and\ \citenamefont
  {Meissner}(2001)}]{Oller:2000fj}%
  \BibitemOpen
  \bibfield  {author} {\bibinfo {author} {\bibfnamefont {J.~A.}\ \bibnamefont
  {Oller}}\ and\ \bibinfo {author} {\bibfnamefont {U.~G.}\ \bibnamefont
  {Meissner}},\ }\bibfield  {title} {\bibinfo {title} {{Chiral dynamics in the
  presence of bound states: Kaon nucleon interactions revisited}},\ }\href
  {https://doi.org/10.1016/S0370-2693(01)00078-8} {\bibfield  {journal}
  {\bibinfo  {journal} {Phys. Lett. B}\ }\textbf {\bibinfo {volume} {500}},\
  \bibinfo {pages} {263} (\bibinfo {year} {2001})},\ \Eprint
  {https://arxiv.org/abs/hep-ph/0011146} {arXiv:hep-ph/0011146} \BibitemShut
  {NoStop}%
\bibitem [{\citenamefont {Au}\ \emph {et~al.}(1987)\citenamefont {Au},
  \citenamefont {Morgan},\ and\ \citenamefont {Pennington}}]{pennington}%
  \BibitemOpen
  \bibfield  {author} {\bibinfo {author} {\bibfnamefont {K.~L.}\ \bibnamefont
  {Au}}, \bibinfo {author} {\bibfnamefont {D.}~\bibnamefont {Morgan}},\ and\
  \bibinfo {author} {\bibfnamefont {M.~R.}\ \bibnamefont {Pennington}},\
  }\bibfield  {title} {\bibinfo {title} {{Meson Dynamics Beyond the Quark
  Model: A Study of Final State Interactions}},\ }\href
  {https://doi.org/10.1103/PhysRevD.35.1633} {\bibfield  {journal} {\bibinfo
  {journal} {Phys. Rev. D}\ }\textbf {\bibinfo {volume} {35}},\ \bibinfo
  {pages} {1633} (\bibinfo {year} {1987})}\BibitemShut {NoStop}%
\bibitem [{\citenamefont {Zhu}\ \emph {et~al.}(2023)\citenamefont {Zhu},
  \citenamefont {Wang}, \citenamefont {Li}, \citenamefont {Wang}, \citenamefont
  {Geng},\ and\ \citenamefont {Xie}}]{Zhu:2022guw}%
  \BibitemOpen
  \bibfield  {author} {\bibinfo {author} {\bibfnamefont {X.}~\bibnamefont
  {Zhu}}, \bibinfo {author} {\bibfnamefont {H.-N.}\ \bibnamefont {Wang}},
  \bibinfo {author} {\bibfnamefont {D.-M.}\ \bibnamefont {Li}}, \bibinfo
  {author} {\bibfnamefont {E.}~\bibnamefont {Wang}}, \bibinfo {author}
  {\bibfnamefont {L.-S.}\ \bibnamefont {Geng}},\ and\ \bibinfo {author}
  {\bibfnamefont {J.-J.}\ \bibnamefont {Xie}},\ }\bibfield  {title} {\bibinfo
  {title} {{Role of scalar mesons a0(980) and a0(1710) in the $D_s^+\to\pi^0
  K^+ K_S^0$ decay}},\ }\href {https://doi.org/10.1103/PhysRevD.107.034001}
  {\bibfield  {journal} {\bibinfo  {journal} {Phys. Rev. D}\ }\textbf {\bibinfo
  {volume} {107}},\ \bibinfo {pages} {034001} (\bibinfo {year} {2023})},\
  \Eprint {https://arxiv.org/abs/2210.12992} {arXiv:2210.12992 [hep-ph]}
  \BibitemShut {NoStop}%
\bibitem [{\citenamefont {del Amo~Sanchez}\ \emph {et~al.}(2010)\citenamefont
  {del Amo~Sanchez} \emph {et~al.}}]{BaBar:2010wfc}%
  \BibitemOpen
  \bibfield  {author} {\bibinfo {author} {\bibfnamefont {P.}~\bibnamefont {del
  Amo~Sanchez}} \emph {et~al.} (\bibinfo {collaboration} {BaBar}),\ }\bibfield
  {title} {\bibinfo {title} {{Evidence for the decay $X(3872) \to \jpsi
  \omega$}},\ }\href {https://doi.org/10.1103/PhysRevD.82.011101} {\bibfield
  {journal} {\bibinfo  {journal} {Phys. Rev. D}\ }\textbf {\bibinfo {volume}
  {82}},\ \bibinfo {pages} {011101} (\bibinfo {year} {2010})},\ \Eprint
  {https://arxiv.org/abs/1005.5190} {arXiv:1005.5190 [hep-ex]} \BibitemShut
  {NoStop}%
\bibitem [{\citenamefont {Lees}\ \emph {et~al.}(2012)\citenamefont {Lees} \emph
  {et~al.}}]{BaBar:2012nxg}%
  \BibitemOpen
  \bibfield  {author} {\bibinfo {author} {\bibfnamefont {J.~P.}\ \bibnamefont
  {Lees}} \emph {et~al.} (\bibinfo {collaboration} {BaBar}),\ }\bibfield
  {title} {\bibinfo {title} {{Study of $X(3915) \to J/\psi \omega$ in
  two-photon collisions}},\ }\href {https://doi.org/10.1103/PhysRevD.86.072002}
  {\bibfield  {journal} {\bibinfo  {journal} {Phys. Rev. D}\ }\textbf {\bibinfo
  {volume} {86}},\ \bibinfo {pages} {072002} (\bibinfo {year} {2012})},\
  \Eprint {https://arxiv.org/abs/1207.2651} {arXiv:1207.2651 [hep-ex]}
  \BibitemShut {NoStop}%
\bibitem [{\citenamefont {Hanhart}\ \emph {et~al.}(2007)\citenamefont
  {Hanhart}, \citenamefont {Kalashnikova}, \citenamefont {Kudryavtsev},\ and\
  \citenamefont {Nefediev}}]{Hanhart:2007yq}%
  \BibitemOpen
  \bibfield  {author} {\bibinfo {author} {\bibfnamefont {C.}~\bibnamefont
  {Hanhart}}, \bibinfo {author} {\bibfnamefont {Y.~S.}\ \bibnamefont
  {Kalashnikova}}, \bibinfo {author} {\bibfnamefont {A.~E.}\ \bibnamefont
  {Kudryavtsev}},\ and\ \bibinfo {author} {\bibfnamefont {A.~V.}\ \bibnamefont
  {Nefediev}},\ }\bibfield  {title} {\bibinfo {title} {{Reconciling the X(3872)
  with the near-threshold enhancement in the D0 anti-D*0 final state}},\ }\href
  {https://doi.org/10.1103/PhysRevD.76.034007} {\bibfield  {journal} {\bibinfo
  {journal} {Phys. Rev. D}\ }\textbf {\bibinfo {volume} {76}},\ \bibinfo
  {pages} {034007} (\bibinfo {year} {2007})},\ \Eprint
  {https://arxiv.org/abs/0704.0605} {arXiv:0704.0605 [hep-ph]} \BibitemShut
  {NoStop}%
\bibitem [{\citenamefont {Zyla}\ \emph {et~al.}(2020)\citenamefont {Zyla} \emph
  {et~al.}}]{Zyla:2020zbs}%
  \BibitemOpen
  \bibfield  {author} {\bibinfo {author} {\bibfnamefont {P.}~\bibnamefont
  {Zyla}} \emph {et~al.} (\bibinfo {collaboration} {Particle Data Group}),\
  }\bibfield  {title} {\bibinfo {title} {{Review of Particle Physics}},\ }\href
  {https://doi.org/10.1093/ptep/ptaa104} {\bibfield  {journal} {\bibinfo
  {journal} {PTEP}\ }\textbf {\bibinfo {volume} {2020}},\ \bibinfo {pages}
  {083C01} (\bibinfo {year} {2020})}\BibitemShut {NoStop}%
\bibitem [{\citenamefont {Flatt\'e}(1976)}]{Flatte:1976xu}%
  \BibitemOpen
  \bibfield  {author} {\bibinfo {author} {\bibfnamefont {S.~M.}\ \bibnamefont
  {Flatt\'e}},\ }\bibfield  {title} {\bibinfo {title} {{Coupled-Channel
  Analysis of the $\pi\eta$ and $K\bar{K}$ Systems Near $K\bar{K}$
  Threshold}},\ }\href {https://doi.org/10.1016/0370-2693(76)90654-7}
  {\bibfield  {journal} {\bibinfo  {journal} {Phys. Lett. B}\ }\textbf
  {\bibinfo {volume} {63}},\ \bibinfo {pages} {224} (\bibinfo {year}
  {1976})}\BibitemShut {NoStop}%
\bibitem [{\citenamefont {Meng}\ \emph {et~al.}(2015)\citenamefont {Meng},
  \citenamefont {Sanz-Cillero}, \citenamefont {Shi}, \citenamefont {Yao},\ and\
  \citenamefont {Zheng}}]{Meng:2014ota}%
  \BibitemOpen
  \bibfield  {author} {\bibinfo {author} {\bibfnamefont {C.}~\bibnamefont
  {Meng}}, \bibinfo {author} {\bibfnamefont {J.~J.}\ \bibnamefont
  {Sanz-Cillero}}, \bibinfo {author} {\bibfnamefont {M.}~\bibnamefont {Shi}},
  \bibinfo {author} {\bibfnamefont {D.-L.}\ \bibnamefont {Yao}},\ and\ \bibinfo
  {author} {\bibfnamefont {H.-Q.}\ \bibnamefont {Zheng}},\ }\bibfield  {title}
  {\bibinfo {title} {{Refined analysis on the X(3872) resonance}},\ }\href
  {https://doi.org/10.1103/PhysRevD.92.034020} {\bibfield  {journal} {\bibinfo
  {journal} {Phys. Rev. D}\ }\textbf {\bibinfo {volume} {92}},\ \bibinfo
  {pages} {034020} (\bibinfo {year} {2015})},\ \Eprint
  {https://arxiv.org/abs/1411.3106} {arXiv:1411.3106 [hep-ph]} \BibitemShut
  {NoStop}%
\bibitem [{\citenamefont {Barnes}\ \emph {et~al.}(2005)\citenamefont {Barnes},
  \citenamefont {Godfrey},\ and\ \citenamefont {Swanson}}]{Barnes:2005pb}%
  \BibitemOpen
  \bibfield  {author} {\bibinfo {author} {\bibfnamefont {T.}~\bibnamefont
  {Barnes}}, \bibinfo {author} {\bibfnamefont {S.}~\bibnamefont {Godfrey}},\
  and\ \bibinfo {author} {\bibfnamefont {E.}~\bibnamefont {Swanson}},\
  }\bibfield  {title} {\bibinfo {title} {{Higher charmonia}},\ }\href
  {https://doi.org/10.1103/PhysRevD.72.054026} {\bibfield  {journal} {\bibinfo
  {journal} {Phys. Rev. D}\ }\textbf {\bibinfo {volume} {72}},\ \bibinfo
  {pages} {054026} (\bibinfo {year} {2005})},\ \Eprint
  {https://arxiv.org/abs/hep-ph/0505002} {arXiv:hep-ph/0505002} \BibitemShut
  {NoStop}%
\bibitem [{\citenamefont {Li}\ \emph {et~al.}(2009)\citenamefont {Li},
  \citenamefont {Meng},\ and\ \citenamefont {Chao}}]{Li:2009ad}%
  \BibitemOpen
  \bibfield  {author} {\bibinfo {author} {\bibfnamefont {B.-Q.}\ \bibnamefont
  {Li}}, \bibinfo {author} {\bibfnamefont {C.}~\bibnamefont {Meng}},\ and\
  \bibinfo {author} {\bibfnamefont {K.-T.}\ \bibnamefont {Chao}},\ }\bibfield
  {title} {\bibinfo {title} {{Coupled-Channel and Screening Effects in
  Charmonium Spectrum}},\ }\href {https://doi.org/10.1103/PhysRevD.80.014012}
  {\bibfield  {journal} {\bibinfo  {journal} {Phys. Rev. D}\ }\textbf {\bibinfo
  {volume} {80}},\ \bibinfo {pages} {014012} (\bibinfo {year} {2009})},\
  \Eprint {https://arxiv.org/abs/0904.4068} {arXiv:0904.4068 [hep-ph]}
  \BibitemShut {NoStop}%
\bibitem [{\citenamefont {Radford}\ and\ \citenamefont
  {Repko}(2007)}]{Radford:2007vd}%
  \BibitemOpen
  \bibfield  {author} {\bibinfo {author} {\bibfnamefont {S.~F.}\ \bibnamefont
  {Radford}}\ and\ \bibinfo {author} {\bibfnamefont {W.~W.}\ \bibnamefont
  {Repko}},\ }\bibfield  {title} {\bibinfo {title} {{Potential model
  calculations and predictions for heavy quarkonium}},\ }\href
  {https://doi.org/10.1103/PhysRevD.75.074031} {\bibfield  {journal} {\bibinfo
  {journal} {Phys. Rev. D}\ }\textbf {\bibinfo {volume} {75}},\ \bibinfo
  {pages} {074031} (\bibinfo {year} {2007})},\ \Eprint
  {https://arxiv.org/abs/hep-ph/0701117} {arXiv:hep-ph/0701117} \BibitemShut
  {NoStop}%
\bibitem [{\citenamefont {Li}\ and\ \citenamefont {Chao}(2009)}]{cc:KTChao}%
  \BibitemOpen
  \bibfield  {author} {\bibinfo {author} {\bibfnamefont {B.-Q.}\ \bibnamefont
  {Li}}\ and\ \bibinfo {author} {\bibfnamefont {K.-T.}\ \bibnamefont {Chao}},\
  }\bibfield  {title} {\bibinfo {title} {{Higher Charmonia and X,Y,Z states
  with Screened Potential}},\ }\href
  {https://doi.org/10.1103/PhysRevD.79.094004} {\bibfield  {journal} {\bibinfo
  {journal} {Phys. Rev. D}\ }\textbf {\bibinfo {volume} {79}},\ \bibinfo
  {pages} {094004} (\bibinfo {year} {2009})},\ \Eprint
  {https://arxiv.org/abs/0903.5506} {arXiv:0903.5506 [hep-ph]} \BibitemShut
  {NoStop}%
\bibitem [{\citenamefont {Wang}\ \emph {et~al.}(2014)\citenamefont {Wang},
  \citenamefont {Yang},\ and\ \citenamefont {Ping}}]{3Pj:Ping}%
  \BibitemOpen
  \bibfield  {author} {\bibinfo {author} {\bibfnamefont {H.}~\bibnamefont
  {Wang}}, \bibinfo {author} {\bibfnamefont {Y.}~\bibnamefont {Yang}},\ and\
  \bibinfo {author} {\bibfnamefont {J.}~\bibnamefont {Ping}},\ }\bibfield
  {title} {\bibinfo {title} {{Strong decays of $\chi_{cJ}(2P)$ and
  $\chi_{cJ}(3P)$}},\ }\href {https://doi.org/10.1140/epja/i2014-14076-y}
  {\bibfield  {journal} {\bibinfo  {journal} {Eur. Phys. J. A}\ }\textbf
  {\bibinfo {volume} {50}},\ \bibinfo {pages} {76} (\bibinfo {year}
  {2014})}\BibitemShut {NoStop}%
\bibitem [{\citenamefont {Guo}\ \emph {et~al.}(2022{\natexlab{b}})\citenamefont
  {Guo}, \citenamefont {Wang}, \citenamefont {Chen},\ and\ \citenamefont
  {Liu}}]{Guo:2022zbc}%
  \BibitemOpen
  \bibfield  {author} {\bibinfo {author} {\bibfnamefont {D.}~\bibnamefont
  {Guo}}, \bibinfo {author} {\bibfnamefont {J.-Z.}\ \bibnamefont {Wang}},
  \bibinfo {author} {\bibfnamefont {D.-Y.}\ \bibnamefont {Chen}},\ and\
  \bibinfo {author} {\bibfnamefont {X.}~\bibnamefont {Liu}},\ }\bibfield
  {title} {\bibinfo {title} {{Connection between near the $D_s^+D_s^-$
  threshold enhancement in $B^+ \to D_s^+D_s^-K^+$ and conventional charmonium
  $\chi_{c0}(2P)$}},\ }\href {https://doi.org/10.1103/PhysRevD.106.094037}
  {\bibfield  {journal} {\bibinfo  {journal} {Phys. Rev. D}\ }\textbf {\bibinfo
  {volume} {106}},\ \bibinfo {pages} {094037} (\bibinfo {year}
  {2022}{\natexlab{b}})},\ \Eprint {https://arxiv.org/abs/2210.16720}
  {arXiv:2210.16720 [hep-ph]} \BibitemShut {NoStop}%
\bibitem [{\citenamefont {Chilikin}\ \emph {et~al.}(2017)\citenamefont
  {Chilikin} \emph {et~al.}}]{Belle:2017egg}%
  \BibitemOpen
  \bibfield  {author} {\bibinfo {author} {\bibfnamefont {K.}~\bibnamefont
  {Chilikin}} \emph {et~al.} (\bibinfo {collaboration} {Belle}),\ }\bibfield
  {title} {\bibinfo {title} {{Observation of an alternative $\chi_{c0}(2P)$
  candidate in $e^+ e^- \rightarrow J/\psi D \bar{D}$}},\ }\href
  {https://doi.org/10.1103/PhysRevD.95.112003} {\bibfield  {journal} {\bibinfo
  {journal} {Phys. Rev. D}\ }\textbf {\bibinfo {volume} {95}},\ \bibinfo
  {pages} {112003} (\bibinfo {year} {2017})},\ \Eprint
  {https://arxiv.org/abs/1704.01872} {arXiv:1704.01872 [hep-ex]} \BibitemShut
  {NoStop}%
\bibitem [{\citenamefont {Deineka}\ \emph {et~al.}(2022)\citenamefont
  {Deineka}, \citenamefont {Danilkin},\ and\ \citenamefont
  {Vanderhaeghen}}]{Deineka:2021aeu}%
  \BibitemOpen
  \bibfield  {author} {\bibinfo {author} {\bibfnamefont {O.}~\bibnamefont
  {Deineka}}, \bibinfo {author} {\bibfnamefont {I.}~\bibnamefont {Danilkin}},\
  and\ \bibinfo {author} {\bibfnamefont {M.}~\bibnamefont {Vanderhaeghen}},\
  }\bibfield  {title} {\bibinfo {title} {{Dispersive analysis of the
  $\gamma\gamma\to D\bar{D}$ data and the confirmation of the $D\bar{D}$ bound
  state}},\ }\href {https://doi.org/10.1016/j.physletb.2022.136982} {\bibfield
  {journal} {\bibinfo  {journal} {Phys. Lett. B}\ }\textbf {\bibinfo {volume}
  {827}},\ \bibinfo {pages} {136982} (\bibinfo {year} {2022})},\ \Eprint
  {https://arxiv.org/abs/2111.15033} {arXiv:2111.15033 [hep-ph]} \BibitemShut
  {NoStop}%
\bibitem [{\citenamefont {Xie}\ \emph {et~al.}(2022)\citenamefont {Xie},
  \citenamefont {Liu},\ and\ \citenamefont {Geng}}]{Xie:2022lyw}%
  \BibitemOpen
  \bibfield  {author} {\bibinfo {author} {\bibfnamefont {J.-M.}\ \bibnamefont
  {Xie}}, \bibinfo {author} {\bibfnamefont {M.-Z.}\ \bibnamefont {Liu}},\ and\
  \bibinfo {author} {\bibfnamefont {L.-S.}\ \bibnamefont {Geng}},\ }\bibfield
  {title} {\bibinfo {title} {{Production rates of $D_{s}^{+}D_{s}^{-}$ and
  $D\bar{D}$ molecules in $B$ decays}},\ }\href@noop {} {\bibfield  {journal}
  {\bibinfo  {journal} {xxx}\ } (\bibinfo {year} {2022})},\ \Eprint
  {https://arxiv.org/abs/2207.12178} {arXiv:2207.12178 [hep-ph]} \BibitemShut
  {NoStop}%
\bibitem [{\citenamefont {Zhou}\ \emph {et~al.}(2015)\citenamefont {Zhou},
  \citenamefont {Xiao},\ and\ \citenamefont {Zhou}}]{Zhou:2015uva}%
  \BibitemOpen
  \bibfield  {author} {\bibinfo {author} {\bibfnamefont {Z.-Y.}\ \bibnamefont
  {Zhou}}, \bibinfo {author} {\bibfnamefont {Z.}~\bibnamefont {Xiao}},\ and\
  \bibinfo {author} {\bibfnamefont {H.-Q.}\ \bibnamefont {Zhou}},\ }\bibfield
  {title} {\bibinfo {title} {{Could the $X(3915)$ and the $X(3930)$ Be the Same
  Tensor State?}},\ }\href {https://doi.org/10.1103/PhysRevLett.115.022001}
  {\bibfield  {journal} {\bibinfo  {journal} {Phys. Rev. Lett.}\ }\textbf
  {\bibinfo {volume} {115}},\ \bibinfo {pages} {022001} (\bibinfo {year}
  {2015})},\ \Eprint {https://arxiv.org/abs/1501.00879} {arXiv:1501.00879
  [hep-ph]} \BibitemShut {NoStop}%
\bibitem [{\citenamefont {Ji}\ \emph {et~al.}(2022{\natexlab{b}})\citenamefont
  {Ji}, \citenamefont {Dong}, \citenamefont {Albaladejo}, \citenamefont {Du},
  \citenamefont {Guo},\ and\ \citenamefont {Nieves}}]{Ji:2022uie}%
  \BibitemOpen
  \bibfield  {author} {\bibinfo {author} {\bibfnamefont {T.}~\bibnamefont
  {Ji}}, \bibinfo {author} {\bibfnamefont {X.-K.}\ \bibnamefont {Dong}},
  \bibinfo {author} {\bibfnamefont {M.}~\bibnamefont {Albaladejo}}, \bibinfo
  {author} {\bibfnamefont {M.-L.}\ \bibnamefont {Du}}, \bibinfo {author}
  {\bibfnamefont {F.-K.}\ \bibnamefont {Guo}},\ and\ \bibinfo {author}
  {\bibfnamefont {J.}~\bibnamefont {Nieves}},\ }\bibfield  {title} {\bibinfo
  {title} {{Establishing the heavy quark spin and light flavor molecular
  multiplets of the X(3872), Zc(3900), and X(3960)}},\ }\href
  {https://doi.org/10.1103/PhysRevD.106.094002} {\bibfield  {journal} {\bibinfo
   {journal} {Phys. Rev. D}\ }\textbf {\bibinfo {volume} {106}},\ \bibinfo
  {pages} {094002} (\bibinfo {year} {2022}{\natexlab{b}})},\ \Eprint
  {https://arxiv.org/abs/2207.08563} {arXiv:2207.08563 [hep-ph]} \BibitemShut
  {NoStop}%
\bibitem [{\citenamefont {Cao}\ \emph {et~al.}(2021{\natexlab{b}})\citenamefont
  {Cao}, \citenamefont {Qi}, \citenamefont {Tang}, \citenamefont {Xue},\ and\
  \citenamefont {Zheng}}]{Cao:2020vab}%
  \BibitemOpen
  \bibfield  {author} {\bibinfo {author} {\bibfnamefont {Q.-F.}\ \bibnamefont
  {Cao}}, \bibinfo {author} {\bibfnamefont {H.-R.}\ \bibnamefont {Qi}},
  \bibinfo {author} {\bibfnamefont {G.-Y.}\ \bibnamefont {Tang}}, \bibinfo
  {author} {\bibfnamefont {Y.-F.}\ \bibnamefont {Xue}},\ and\ \bibinfo {author}
  {\bibfnamefont {H.-Q.}\ \bibnamefont {Zheng}},\ }\bibfield  {title} {\bibinfo
  {title} {{On leptonic width of $X(4260)$}},\ }\href
  {https://doi.org/10.1140/epjc/s10052-020-08813-y} {\bibfield  {journal}
  {\bibinfo  {journal} {Eur. Phys. J. C}\ }\textbf {\bibinfo {volume} {81}},\
  \bibinfo {pages} {83} (\bibinfo {year} {2021}{\natexlab{b}})},\ \Eprint
  {https://arxiv.org/abs/2002.05641} {arXiv:2002.05641 [hep-ph]} \BibitemShut
  {NoStop}%
\end{thebibliography}%

\end{document}